\documentclass[manuscript,nonacm,acmsmall,screen]{acmart}

\usepackage[LGR,T1]{fontenc}

\usepackage{float}
\usepackage{pgf}
\usepackage{pgfplots,pgfplotstable}
\usepackage{tikz}
\usetikzlibrary{shapes,backgrounds}
\usetikzlibrary{arrows,shapes,fit,automata,positioning,decorations,calc}
\usetikzlibrary{spy,backgrounds}
\usepackage{siunitx}
\pgfplotsset{compat=1.12}
\usepackage{tabulary}
\usepackage{tabularx}
\usepackage{multirow}
\usepackage{booktabs}
\usepackage{listings}
\usepackage{acro}
\usepackage{tcolorbox}
\usepackage{amsmath}
\usepackage{newtxmath}
\let\savedbaselinestretch\baselinestretch
\usepackage{oz}
\let\baselinestretch\savedbaselinestretch

\makeatletter
\def\@setfontsize#1#2#3{%
  \@nomath#1%
  \ifx\protect\@typeset@protect
    \let\@currsize#1%
  \fi
  \fontsize{#2}{#3}\selectfont
  \ifx\baselinestretch\@empty
    \setlength{\zedbaselineskip}{\baselineskip}%
  \else
    \setlength{\zedbaselineskip}{\baselineskip/\real{\baselinestretch}}%
  \fi
  \zedbaselineskip\zedbaselinestretch\zedbaselineskip
  \sbox\zstrutbox{%
    \vrule height.7\zedbaselineskip depth.3\zedbaselineskip width\z@
  }%
}
\makeatother
\usepackage{mathtools}
\usepackage{array}
\usepackage{url}
\usepackage{etoolbox}
\usepackage[utf8x]{inputenc}  
\usepackage[T1]{fontenc}
\usepackage{lmodern}
\usepackage{graphicx}
\usepackage{paralist}
\usepackage{wrapfig}
\usepackage{lipsum}  
\usepackage{microtype}

\usepackage{cleveref}
\usepackage{titlesec}

\usepackage{amsfonts} %
\usepackage{calligra} %

\definecolor{graphFirst}{RGB}{2,136,209} 
\definecolor{graphSecond}{RGB}{211,47,47} 
\definecolor{graphThird}{RGB}{245,124,0} 
\definecolor{graphFourth}{RGB}{56,142,60} 
\definecolor{graphFifth}{RGB}{81,45,168} 
\definecolor{graphSixth}{RGB}{69,90,100} 
\definecolor{graphSeventh}{RGB}{251,192,45} 

\definecolor{backgroundFirst}{RGB}{129,212,250} 
\definecolor{backgroundSecond}{RGB}{239,154,154} 
\definecolor{backgroundThird}{RGB}{255,204,128} 
\definecolor{backgroundFourth}{RGB}{165,214,167} 
\definecolor{backgroundFifth}{RGB}{179,157,219} 
\definecolor{backgroundSixth}{RGB}{176,190,197} 
\definecolor{backgroundSeventh}{RGB}{255,245,157} 

\lstdefinestyle{c}{
	language = C,
	commentstyle = \color{graphFourth},
	keywordstyle = \color{graphFirst}\bfseries,
	stringstyle = \color{graphSecond},
}
\lstdefinestyle{timetide}{
	language = Timetide,
	commentstyle = \color{graphFourth},
	keywordstyle = \color{graphFirst}\bfseries,
	stringstyle = \color{graphSecond},
}
\lstdefinestyle{esterel}{
	language = Esterel,
	commentstyle = \color{graphFourth},
	keywordstyle = \color{graphFirst}\bfseries,
	stringstyle = \color{graphSecond},
}
\lstdefinestyle{lf}{
	language = LINGUAFRANCA,
	commentstyle = \color{graphFourth},
	keywordstyle = \color{graphFirst}\bfseries,
	stringstyle = \color{graphSecond},
}

\lstdefinelanguage{Esterel}{
    morekeywords={abort, and, await, call, case, combine, constant do, each,
                  else, elsif, emit, end, every, exec, exit, false, function,
                  halt, handle, if, immediate, in, input, inputoutput, loop,
                  mod, module, not, nothing, or,output, pause, positive, pre,
                  present, procedure, relation, repeat, return, run, sensor,
                  signal, suspend, sustain, task, then, tick, timeout, times,
                  trap, true, type, upto, var, watching, weak, when, with
                 },
    sensitive=true,
    morecomment=[l]{\%},
    morestring=[b]",
}

\lstdefinelanguage{LINGUAFRANCA}{
    morekeywords={reactor, reaction, timer, new, after, input, output, state, if
                 },
    sensitive=true,
    morecomment=[l]{\%},
    morestring=[b]",
}

\lstdefinelanguage{Timetide}{
    morekeywords={abort, and, delay, await, call, case, combine, constant do, each, chan, send, receive, sync, period, duration, offset, channel, ,
                  else, elsif, emit, end, every, exec, exit, false, function,
                  halt, handle, if, immediate, in, input, inputoutput, loop,
                  mod, module, not, nothing, or,output, pause, positive, pre,
                  present, procedure, relation, repeat, return, run, sensor,
                  signal, suspend, sustain, task, then, localtick, timeout, times, const, pareach, foreach,
                  trap, true, type, upto, var, watching, weak, when, with, read, <>, toplevel, preamble, fresh,
                 },
    otherkeywords = {<>},
    sensitive=true,
    morecomment=[l]{\%},
    morestring=[b]",
}
\newcolumntype{P}[1]{>{\raggedright\arraybackslash}p{#1}<{\hspace{0pt}}}
\tcbset{
    colframe=black,
    colback=white,
    sharp corners,
    boxrule=0.5mm,
    left=2pt,
    right=2pt,
    top=1pt,
    bottom=1pt,
    boxsep=0pt 
}
\lstdefinelanguage{LTL}{
    morekeywords={ltl},
    sensitive=true,
    morecomment=[l]{\/\/},
    morestring=[b]",
}

\let\origthelstnumber\thelstnumber
\makeatletter
\newcommand*\Suppressnumber{%
	\lst@AddToHook{OnNewLine}{%
		\let\thelstnumber\relax%
		\advance\c@lstnumber-\@ne\relax%
	}%
}

\newcommand*\Reactivatenumber{%
	\lst@AddToHook{OnNewLine}{%
		\let\thelstnumber\origthelstnumber%
		\advance\c@lstnumber\@ne\relax}%
}

\usepackage{algorithmic}
\usepackage{fancybox}
\usepackage{caption}
\usepackage{subcaption}
\usepackage{bm}
\DeclareCaptionType{equ}[][]

\usepackage{etoolbox}

\usepackage{thmtools,blindtext}
 
\declaretheorem{lemma} 
\declaretheorem{definition}
\declaretheorem[name={Theorem}]{thm}
\declaretheoremstyle[%
	spacebelow=10pt,
	qed=\qed%
]{mystyle} 
\declaretheorem[name={Proof},style=mystyle,unnumbered]{pf}

\newcommand{\xrsquigarrow}[2]{%
  \mathrel{\mathop{\rightsquigarrow}\limits^{#1}_{#2}}%
}

\newcommand{\change}[1]{{#1}}

\newcommand{\remove}[1]{}
\newcommand{\ignore}[1]{}

\newcommand{\squishlist}{
 \begin{list}{-}
  { \setlength{\itemsep}{0pt}
     \setlength{\parsep}{1pt}
     \setlength{\topsep}{1pt}
     \setlength{\partopsep}{0pt}
     \setlength{\leftmargin}{0.9em}
     \setlength{\labelwidth}{1.5em}
     \setlength{\labelsep}{0.4em} } }
\newcommand{\squishend}{
  \end{list}  }

\newcommand{\Tau}{{\mathcal{T}}}

\makeatletter
\ifx\proof\undefined
\newenvironment{proof}[1][\protect\proofname]{\par
\normalfont\topsep6\p@\@plus6\p@\relax
\trivlist
\itemindent\parindent
\item[\hskip\labelsep\scshape #1]\ignorespaces
}{%
\endtrivlist\@endpefalse
}
\providecommand{\proofname}{Proof}
\fi

\makeatother

\DeclareAcronym{FIFO}{short=FIFO, long= First In First Out}
\DeclareAcronym{KPN}{short=KPN, long= Kahn Process Network}
\DeclareAcronym{LTTA}{short=LTTA, long=Loosely Time-Triggered Architecture, short-indefinite=an}
\DeclareAcronym{LF}{short=LF, long= Lingua Franca, short-indefinite=an, first-style=long, subsequent-style=long}
\DeclareAcronym{TTP}{short=TTP, long= Time Triggered Protocol}
\DeclareAcronym{NTP}{short=NTP, long= Network Time Protocol, short-indefinite=an}
\DeclareAcronym{PTP}{short=PTP, long= Precision Time Protocol}
\DeclareAcronym{TT}{short=TT, long= Timetide, first-style=long, subsequent-style=long}
\DeclareAcronym{LET}{short=LET, long= Logical Execution Time}
\DeclareAcronym{LSN}{short=LSN, long= Logical Synchrony Network, short-indefinite=an}
\DeclareAcronym{FFP}{short=FFP, long= Finite FIFO Platform, short-indefinite=an}
\DeclareAcronym{WCET}{short=WCET, long= Worst Case Execution Time}
\DeclareAcronym{sLET}{short=sLET, long= Synchronous Logical Execution Time, short-indefinite=an}
\DeclareAcronym{SL-LET}{short=SL-LET, long=System-Level Logical Execution Time, short-indefinite=an}
\DeclareAcronym{MoC}{short=MoC, long=Model of Computation, short-indefinite=an}

\DeclareAcronym{TDMA}{
    short=TDMA,
    long=Time Division Multiple Access
}
\DeclareAcronym{LS}{short=LS, long= logical synchrony, short-indefinite=an, first-style=long, subsequent-style=long}
\DeclareAcronym{RSU}{short=RSU, long=Roadside Unit, short-indefinite=an}
\DeclareAcronym{SOS}{short=SOS, long=Structural Operational Semantics, short-indefinite=an}

\newcommand{\bittide}{{bittide}}

\DeclareAcronym{LTL}{
	short=LTL,
	long=Linear Temporal Logic,
	short-indefinite=an,
}
\DeclareAcronym{CBMC}{
	short=CBMC,
	long=C Bounded Model Checker
}

\DeclareAcronym{ILP}{
	short=ILP,
	long=Integer Linear Programming,
	short-indefinite=an,
	long-indefinite=an
}

\DeclareAcronym{SA}{
	short=SA,
	long=Simulated Annealing,
	short-indefinite=an
}

\DeclareAcronym{WCRT}{
	short=WCRT,
	long=Worse Case Reaction Time
}

\DeclareAcronym{LCM}{
	short=LCM,
	long=Least Common Multiple,
	short-indefinite=an
}

\usepackage{xspace}
\usepackage{fontawesome}

\begin{document}

\title{Timetide: A programming model for logically synchronous distributed systems}

\author{Logan Kenwright}
\email{logan.kenwright@auckland.ac.nz}
\orcid{0000-0003-0923-0307}
\affiliation{%
  \institution{University of Auckland}
  \city{Auckland}
  \country{New Zealand}
}

\author{Partha Roop}
\email{p.roop@auckland.ac.nz}
\orcid{0000-0001-9654-5678}
\affiliation{%
  \institution{University of Auckland}
  \city{Auckland}
  \country{New Zealand}
}

\author{Nathan Allen}
\email{nathan.allen@aut.ac.nz}
\orcid{0000-0001-7876-819X}
\affiliation{%
  \institution{Auckland University of Technology}
  \city{Auckland}
  \country{New Zealand}
}

\author{C\u{a}lin Ca\c{s}caval}
\email{cascaval@google.com}
\orcid{0000-0002-2780-6763}
\affiliation{%
  \institution{Google Deepmind}
  \city{Auckland}
  \country{New Zealand}
}

\author{Avinash Malik}
\email{avinash.malik@auckland.ac.nz}
\orcid{0000-0002-7524-8292}
\affiliation{%
  \institution{University of Auckland}
  \city{Auckland}
  \country{New Zealand}
}

\renewcommand{\shortauthors}{Kenwright et al.}





\begin{abstract}
    Massive strides in deterministic models have been made
    using synchronous languages. They are mainly focused on centralised
    applications, as the traditional approach is to compile away the concurrency.
    Time triggered languages such as Giotto and
    \acf{LF} are suitable for distribution albeit that they rely on expensive
   physical clock synchronisation, which is both expensive and may suffer from scalability.
    Hence, deterministic programming of distributed systems remains challenging.

    We address the challenges of deterministic distribution by
    developing a novel multiclock semantics of synchronous programs.  The developed semantics is
    amenable to seamless distribution. Moreover, our programming model, \acf{TT},  alleviates the
    need for physical clock synchronisation by building on the recently proposed \emph{logical
    synchrony} model for distributed systems.
    We discuss the important aspects of distributing
    computation, such as network communication delays, and explore
    the formal verification of \ac{TT} programs. To the best of our knowledge,
    \ac{TT} is the first multiclock synchronous language that is both amenable
    to distribution and formal verification without the need for
    physical clock synchronisation or clock gating.

\end{abstract}

\maketitle

\section{Introduction}
Deterministic programming of distributed systems remains challenging~\cite{lee2021determinism} in spite of decades of formal models for such systems, such as \acp{KPN}~\cite{gilles1974semantics}.
Despite the advantages of deterministic execution, the vast majority of distributed systems are asynchronous and thus non-deterministic. For example, the widely used
actor-based models are inherently
non-deterministic~\cite{lohstroh2021toward}. Non-deterministic
concurrency, while being a desirable feature for
specification~\cite{hoare1985communicating, milner1984lectures}, is hard to
verify and debug. In contrast, synchronous languages~\cite{benveniste2003synchronous} alleviate this by ``compiling away'' the concurrency, eliminating any run-time uncertainty. However, the distribution of these programs remains challenging. While many approaches have been studied~\cite{girault2005survey}, they are not scalable in general. Hence, achieving both good performance and determinism is an unsolved problem, as most approaches rely on expensive physical clock synchronisation.

The need to synchronise distributed components in a system raises at least two major challenges. One is achieving maximum throughput --- when a system has dependencies between components that must be synchronised on, these cannot execute freely. Rather, they must wait for any upstream data to arrive. Consequently, systems which execute in lock-step, such as \acp{LTTA}~\cite{tripakis2008implementing}, experience worsening throughput proportional to the transmission delay between machines. The second challenge is balancing the requirements of consistency and availability~\cite{lee2023consistency}. In a distributed system, a designer must choose between the ability to retrieve the true value of a shared variable (consistency) and the amount of time it takes a system to respond to a request (availability). Typically, synchronous systems are completely consistent, which comes at the cost of availability. By specifying delays between tasks, we propose to relax the consistency requirement, and thus improve the availability of the system. The recently proposed \emph{\ac{LS}} model~\cite{lall_logical_synchrony_2024} describes an abstract model for distributed systems where delays are exactly specified, and forms the basis for the \ac{TT} language presented in this work.

\subsection{Logical Synchrony}
While there exist many approaches
methods~\cite{corbett_spanner_2013,li_sundial_2020} that exchange timestamps based on protocols such as \ac{PTP}~\cite{ptp}, these are expensive to build for high precision accuracy. 
\emph{\Ac{LS}}~\cite{lall_logical_synchrony_2024} is a viable alternative to physical clock synchronization. \Ac{LS} provides a shared notion of time sufficient for reasoning about causality without requiring a shared system-wide clock. Applications running on the system use a local logical clock and derived knowledge of their peers' logical clocks to coordinate their actions, which replaces the need to reference physical time.

\change{
	\Acp{LSN}~\cite{kenwright2024logical} are a graph-based \ac{MoC} based on logical synchrony, which capture computational machines as the nodes of the graph and uni-directional links between them for communication.
	Each link has an invariant logical delay between production and consumption of tokens. 
	This ensures communication-level determinism, as proven in \cite{kenwright2024logical}, by 
	establishing that the execution forms a Complete Partial Order. We recap the definition of \ac{LSN} from \cite{kenwright2024logical} for completeness.
	
	\begin{definition}
		\Iac{LSN} is defined as a tuple $\mathcal{L} = (G, \Theta, \lambda)$, where:
		\begin{itemize}
			\item $G = (V, E)$ is a directed graph of $V$ vertices (the distributed machines $\mathcal{M}$) and $E \subseteq V \times V$ edges connecting them (the communication links), with $\forall_{(v1, v2) \in E} : v1 \neq v2$,
			\item $\Theta$ represents the set of local clock valuations for each node, such that the valuation of a clock $\theta_i \in \Theta$ for machine $\mathcal{M}_i \in V$ at a given time $t$ is denoted as $\theta_i(t) \in \mathbb{N}$, and
			\item $\lambda : E \rightarrow \mathbb{Z}$ captures the communication delay for each link in the network.
			For simplicity, the delay between $\mathcal{M}_i$ and $\mathcal{M}_j$ is denoted as $\lambda_{i \rightarrow j} = \lambda((\mathcal{M}_i, \mathcal{M}_j))$.
		\end{itemize}
	\end{definition}

}

\begin{figure}[htbp]
	\begin{center}
		\includegraphics[clip=true,trim=0.1cm 0.05cm 0.2cm 0.025cm,width=0.3\linewidth]{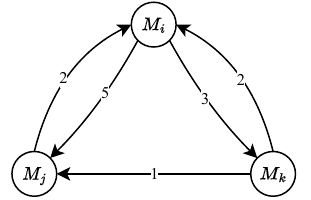}
	\end{center}
	\caption{An example of \iacl*{LSN}}
	\label{fig:lsn}
\end{figure}

\Cref{fig:lsn} shows an example of \iac{LSN}\change{, where the numbers on each edge correspond to the logical delays $\lambda$}.
At each logical tick for a single node, a unit of data called a \textit{frame} is consumed on each incoming edge of the graph, and a frame is produced on each outgoing edge.
A send event at a node is delayed by a fixed logical delay before it is consumed by a receiving node.
This relationship is captured in \Cref{eqn:logicaldelay} as follows:
\begin{equation}{\theta_j(t_{receive}) = \theta_i(t_{send}) + \lambda_{i\rightarrow{}j}} \label{eqn:logicaldelay}
\end{equation}
where $\theta_i(t_{send})$ is the logical clock value of a sender at the time it sends, $\lambda_{i\rightarrow{}j}$ is the fixed logical latency between the sender and receiver, and $\theta_j(t_{receive})$ is the logical clock value of the receiver at the time it receives the message. Notably, the specific values of $t_{send}$ and $t_{receive}$ are not important, as the logical delay is fixed.

\Ac{LS} may be implemented at the system level using the approach given by the \emph{\bittide{}}~\cite{lall2022modeling} protocol, where nodes synchronize by monitoring the rate of communication with their neighbors without requiring a global clock. Alternatively, \ac{LS} has also been implemented using Kahn-like token pushing networks~\cite{kenwright2024logical}. In both approaches, when viewed from the outside, the shared logical time is fully disconnected from physical wall-clock time, meaning that logical time steps can vary in physical duration.
\change{There is therefore no requirement for systems to be completely synchronised in their logical clocks at any physical instant from the point of view of a hypothetical omniscient observer, as long as the logical delay invariance can always be maintained.}
\change{The} \ac{LS} model only describes the communication behaviour and does not provide a programming model for the tasks running on the nodes.
\change{We will elaborate on this in the following sections.}


\subsection{\Acl*{LET} Task Model}

\begin{figure}[htbp]
	\begin{center}
		\includegraphics[clip=true,trim=0.3cm 0.26cm 0.24cm 0.5cm,width=0.5\linewidth]{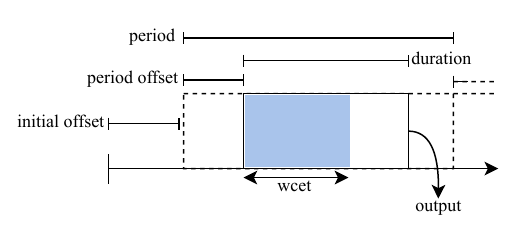}
	\end{center}
	\caption{Structure of a (typical) \acs*{LET} task}
	\label{fig:let_task}
\end{figure}

The \ac{LET} model is a programming abstraction which describes networks of communicating tasks.
\Ac{LET} is commonly used for modelling timed concurrent systems as it exhibits timing-determinism. Such systems are typically cyber-physical systems, composed of one or more processing cores driven by a single clock.
\Iac{LET} task, shown in \Cref{fig:let_task}, is expressed in languages like Giotto~\cite{kirsch2012logical}.
Each task has a \emph{period} describing how often a task begins execution, a \emph{duration} which describes how long between the time a task begins until it emits an output, an \emph{initial offset} describing the time between the start of synchronous clock and the first release of the task for execution, and a \emph{period offset} describing how far into the repetition cycle the task body begins.

As an alternative, the \ac{LET} model may be combined with synchronous programming in the \ac{sLET} model~\cite{siron2022synchronous}. Rather than synchronise on physical timestamps, \ac{sLET} tasks synchronise on named clocks, gaining the benefits of synchronous programming, but cannot express large transmission delays.
The \ac{SL-LET} model~\cite{ernst2018system} explicitly specifies transmission delays, allowing for the modelling of systems with non-negligible latency. Unlike the \ac{sLET} model, the \ac{SL-LET} model does not make use of logical time for verification.

\subsection{\change{\Ac{TT}}: deterministic distributed programming}

\change{Currently, there are no programming models which combine the purely synchronous approach of \ac{sLET} with the explicit communication delays of \ac{SL-LET}.
Existing distributed programming languages generally ignore latency and require that a designer chooses between non-determinism or expensive physical clock synchronisation. 
\Ac{TT} is proposed to unify the \ac{sLET} and \ac{SL-LET} approaches by allowing the specification of logical communication delays between tasks.
This allows the user to form powerful execution pipelines.

Thus, for the first time, we can express high-performance distributed systems with deterministic execution and no physical clock synchronisation in the presence of non-negligible transmission latencies.
By leveraging \acp{LSN} for the underlying execution model, we separate the synchronisation layer from the application layer and allow \ac{TT} programs to execute over a wide range of synchronisation methods.}

The main contributions of \ac{TT} are as follows:
\begin{enumerate}
  \item We introduce the \ac{TT} language for deterministic distributed systems. \change{\Ac{TT}} is the first language based on logical synchrony~\change{\cite{lall_logical_synchrony_2024}}.
  \item \Ac{TT} treats communication delays as first-class citizens, a feature absent in other languages \change{for} distributed systems.
  \item We \change{introduce} \ac{LSN}-compatibility to specify architectures that can deploy \ac{TT} \change{programs}, \change{executing either as centralised or distributed applications}.
  \item \change{We present the formal semantics for \ac{TT}. \change{This facilitates} deterministic distribution \change{without the need for} physical clock synchronisation.}
\end{enumerate}

The paper is structured as follows.
Firstly, \change{we} introduce a motivating example in \Cref{sec:app_examples}. Subsequently, we present the constructs of \ac{TT} in \Cref{sec:syntax}. In \Cref{sec:semantics}, we propose a set of statements, we term the \emph{kernel language}, using which we can express all other
language constructs. We use these statements to develop the operational semantics of \ac{TT} and formalise the key properties of the language.
\change{Subsequently, in \Cref{sec:determinism}, we formally prove the determinism of this language.}
In \Cref{sec:verification} we develop a simple source to source translation to the well-known Esterel language, to leverage existing tools for compilation and verification.
In \Cref{sec:mapping}, we illustrate how \ac{TT} programs are implementable over \ac{LSN}-compatible architectures. In \Cref{sec:comparison}, we compare and benchmark \ac{TT} against the deterministic language \textit{Lingua Franca}.
The paper finishes by comparing this approach to the related work and making concluding remarks, including the scope for future developments
of \ac{TT}.


\section{Motivating Example}
\label{sec:app_examples}
\begin{figure}[htbp]
	\begin{minipage}[c]{.47\textwidth}
		\centering
		\footnotesize
		\captionof{table}{A price spread of the three best bid and ask prices}
		\begin{tabular}{|c|c|c|}
			
			\hline
			\text{Side} & \text{Price} & \text{Quantity} \\
			\hline
			\text{Buy} & 100 & 97 \\
			\text{Buy} & 99 & 32 \\
			\text{Buy} & 98 & 121 \\
			\text{Sell} & 101 & 88 \\
			\text{Sell} & 102 & 42 \\
			\text{Sell} & 103 & 79 \\
			\hline
		\end{tabular}
		\label{table:price_spread}
	\end{minipage}
	\hfill
	\begin{minipage}[c]{.47\textwidth}
		\begin{center}
			\includegraphics[clip=true,trim=0cm 0cm 0cm 0cm,width=\linewidth]{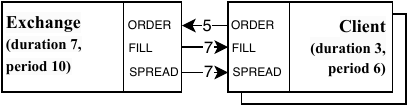}
			\captionof{figure}{A financial trading system modelled in \acl*{TT} \change{with logical delays marked on edges}}
			\label{fig:financial_trading}
		\end{center}
	\end{minipage}
\end{figure}

We motivate \ac{TT} using an application modelling a financial trading system, where logical time, not physical time, is used to arbitrate trades, ensuring that all traders get a fair chance to participate.
The system consists of a single
\textit{Exchange}, and an arbitrary number of \textit{Traders}. Both the traders and the exchange execute periodic tasks. Periodically, the exchange performs an execution where it checks whether any trade orders have arrived, matches the buy/sell orders, and sends order confirmations back to the traders, as well as the price spread showing the best buy/sell prices. The price spread is the best bid and ask prices, along with the quantity available at each price, as shown in \Cref{table:price_spread}.

The traders similarly perform a periodic task, albeit at a different rate. Each cycle a trader reads any responses to their previous orders and the latest price spread. If a new spread has been received since the last cycle the trader may decide to place a new order. Each task takes a number of ticks to complete before emitting a value. This program is visualised in \Cref{fig:financial_trading}, showing the chosen durations, periods, and transmission delays.

In a practical scenario the values for the periods, durations, and delays will be chosen based on the system specifications. For example, in a trading setting, enforcing the delay between each of the traders and the exchange to be equal ensures a fair playing field. Alternatively, the exchange may choose to create tiers of service that traders can subscribe to.

\section{The Timetide language}
The \ac{TT} language is designed to support the \ac{LS} model abstractions: a network of communicating tasks, synchronized on a logical clock. In \ac{TT} the fundamental unit of modularity is the \textit{module}, which can either perform a periodic {\em task} or instantiate other modules. Thus, any module may be chosen to be the top level.
Tasks follow the LET model: they have periodic release times (ticks at which they begin a cycle of execution) and a duration (or deadline, by which computation must end). Outputs are ready to send once the deadline has passed. Tasks communicate using {\em channels} with fixed logical delays. Channels are read from and written to on each clock tick, however tasks only sample the most recent value read from the channel at their release time, which may cause interim values to be missed. Yet, because deadlines and delays are exact, programs can be analyzed to detect such cases. Similarly, outputs are only updated at the end of a task duration.



\subsection{Syntax}
\label{sec:syntax}

The \ac{TT} syntax is inspired by Esterel, since it provides a solid foundation for expressing synchrony. Moreover, as we will show later, we can translate \ac{TT} to Esterel and take advantage of its verification tools. The list of \ac{TT} statements is shown in \Cref{tbl:kernel}.

\begin{table}[htbp]
	\centering
	\caption{\Acl*{TT} statements}
	\small
	\begin{tabular}{p{0.49\linewidth}p{0.5\linewidth}}
	\hline

	\textbf{Statement}                   & \textbf{Meaning} \\
	\hline
	
		{\tt\footnotesize module \emph{m}: ... end module} & declare a module \\
		{\tt\footnotesize input [const] <name> : \emph{type}} & {declare an input port to a module} \\
		{\tt\footnotesize output <name> : \emph{type}} & {declare an output port to a module} \\
		{\tt\footnotesize channel \emph{ch} : {[}\emph{type}{]} delay \emph{$\delta$}}  & declare a named channel {\tt\footnotesize\em{ch}} of {\tt\footnotesize\em{type}} with delay $\delta$\\
		{\tt\footnotesize\emph{t}$\mathbf{\boldsymbol{<>}}$\emph{u}} & concurrently execute program statements {\tt\footnotesize\em{t}} and
		{\tt\footnotesize\em{u}}\\
		{\tt\footnotesize run <module> [<channel>/<port>, ...]} & instantiate a module with channel bindings \\
		{\tt\footnotesize foreach i in <const or num> \{t\}} & execute {\tt\footnotesize\em{t}} in sequence for each value of {\tt\footnotesize\em{i}} \\ 
		{\tt\footnotesize pareach i in <const or num> \{t\}} & execute {\tt\footnotesize\em{t}} in a parallel thread for each value of {\tt\footnotesize\em{i}} \\ 
		{\tt\footnotesize var \emph{v} : \emph{type} [ = <initial>] in \emph{t} end}  & declare a local variable {\tt\footnotesize\em{v}} of {\tt\footnotesize\em{type}} scoped to {\tt\footnotesize\em{t}} \\
		{\tt\footnotesize const \emph{c} : \emph{type} = <value>} & declare a compile-time constant \\
		
		{\tt\footnotesize\emph{t};\emph{u}}  & run {\tt\footnotesize\em{t}}, and then
			{\tt\footnotesize\em{u}} in sequence \\
		{\tt\footnotesize task(period=\emph{p},duration=\emph{d},offset=\emph{o}): \emph{t} end} & run {\tt\footnotesize\em{t}} every {\tt\footnotesize\em{p}} ticks, for {\tt\footnotesize\em{d}} ticks, starting at {\tt\footnotesize\em{o}} ticks \\
    {\tt\footnotesize [weak?] abort t when [immediate?] \emph{expr}} & (Weak) abort the body when {\tt\footnotesize\em{expr}} becomes true\\

            $v = f(...)$                          & assign $v$ with the expression $f$                                 \\

            {\tt\footnotesize if} $c(...)$ {\tt\footnotesize {\emph{t}} else {\emph{u}}}
                                                   & run {\tt\footnotesize\em{t}} if condition
            $c$; otherwise {\tt\footnotesize\em{u}}                                                                         \\
  	{\tt\footnotesize <expr>} & evaluate an expression \\
	{\tt\footnotesize send \emph{ch(<expr>)}} & send a value along channel {\tt\footnotesize\em{ch}}\\
  	{\tt\footnotesize fresh(\emph{ch})} & true if the value in {\tt\footnotesize\em{ch}} has not been sampled yet \\
	  \texttt{\footnotesize+}, \texttt{\footnotesize-}, \texttt{\footnotesize/}, \texttt{\footnotesize*}, \texttt{\footnotesize>}, \texttt{\footnotesize<}, \texttt{\footnotesize<=}, \texttt{\footnotesize>=}, \texttt{\footnotesize!} & arithmetic operators\\
	  \texttt{\footnotesize{and}},\texttt{\footnotesize{or}} & boolean operators \\
	  \texttt{\footnotesize<ident>} & read a variable, constant, or sampled channel \\
		\hline
      \end{tabular}
      \label{tbl:kernel}
\end{table}

We illustrate the syntax using an implementation of the financial trading example from \Cref{sec:app_examples}.
\Cref{fig:timetide_exchange_top} shows the top level \ac{TT} specification of the financial trading system example which instantiates a single exchange \textit{center} and two \textit{traders}. Channels are specified with their types and logical delays. Each input and output port of a module is mapped by the top level to the fixed-delay channels in the network. Threads are specified through parallel arms of the parallel \texttt{<>} operator. Threads will run concurrently, co-located or distributed based on a schedule defined by the mapping to an actual architecture. Channel routing is automatically inferred based on their usage in the module, in this case by the \texttt{run} statement which instantiates a module template.

Two types of basic iterators are supported in \acl{TT}: \texttt{foreach}, which declares a sequential loop, and \texttt{pareach}, which declares a (distributed) parallel loop. Both of these are simple substitution macros which unravel into a sequence (\texttt{t;u;v}) and parallel blocks (\texttt{t<>u<>v}) respectively. Only a compile-time constant is allowed as the iterator bound, as all threads in \acl{TT} have static lifetime.

\Cref{fig:timetide_exchange_center} shows the declaration of the module for the exchange Center. \Cref{line:tt_task:sugar:module} assigns a module name. As we've seen, the module name is then used by the \texttt{run} statement to spawn work. Each module may have input and output declarations (\cref{line:tt_task:sugar:input} - \ref{line:tt_task:sugar:output}), which may be endpoints of a channel or a compile-time constant which is passed in.
The body of the module consists of statements that are evaluated sequentially. The \texttt{var} statement on \cref{line:tt_task:sugar:var} declares a scoped variable block. Variables are mutable values which are local to their enclosing scope and cannot be shared between threads.
The \texttt{task} statement (\cref{line:tt_task:sugar:task}) declares a repeating periodic task (similar to a typical \ac{LET} task), specifying its period, duration, and optional offset from the beginning of the period to start of work. \Cref{line:tt_task:sugar:receive} shows a variable assignment, and the current value of the \texttt{order} channel being read. The \texttt{send} statement (\cref{line:tt_task:sugar:send}) writes to a buffer in data memory for the associated channel such that at the end of the task body the value is sent to the tail of channel. We also show the \ac{TT} specifications for the trader component in \Cref{fig:timetide_exchange_trader}.

\begin{figure}[htbp]
	\begin{lstlisting}[style=timetide]
module toplevel:
	const TRADERS : int = 2;
	channel orders : Order[TRADERS] delay 5;
	channel fills : MatchedOrders[TRADERS] delay 7;
	channel spreads : Spread[TRADERS] delay 7;
	{
		run Center(TRADERS/TRADERS, orders/orders, fills/fills, spreads/spreads);
	}
	<>
	pareach i in TRADERS {
		run Trader(spreads[i]/price_spread, orders[i]/order, fills[i]/fill, i/id);
	}
end module;
	\end{lstlisting}
	\caption{The \acl*{TT} code of the top level module}
	\label{fig:timetide_exchange_top}
\end{figure}

\begin{figure}[htbp]
	\begin{lstlisting}[style=timetide]
module Center: (*@\label{line:tt_task:sugar:module}@*)
	input const TRADERS : int;
	input orders : Order[TRADERS]; (*@\label{line:tt_task:sugar:input}@*)
	output fills : MatchedOrders[TRADERS];
	output spreads : Spread[TRADERS]; (*@\label{line:tt_task:sugar:output}@*)
	var orderbook : OrderBook = create_orderbook(4) in (*@\label{line:tt_task:sugar:var}@*)
		task(period=10, duration=7): (*@\label{line:tt_task:sugar:task}@*)
			foreach i in TRADERS {
				if (fresh(order[i])) {
					orderbook = insert_order(orderbook, order[0]); (*@\label{line:tt_task:sugar:receive}@*)
				}
			}
			var matched_orders : MatchedOrders = run_matching(orderbook) in
				send fill(matched_orders); (*@\label{line:tt_task:sugar:send}@*)
			end var;
			var price_spread : Spread = get_spread(orderbook) in
				send fill(price_spread);
			end var;
		end task;
	end var;
end module;
	\end{lstlisting}
	\caption{The \acl*{TT} code of the controller module}
	\label{fig:timetide_exchange_center}
\end{figure}

Parallel instantiations of modules communicate over channels --- point-to-point communication media that behave as \ac{FIFO} queues. At each local tick on the sender's clock, a value is sent on the channel, and at each local tick on the receiver's clock, a value is read from the channel.
The value sent on the channel is specified by the most recent \texttt{send} statement used within the task body. If this tick does not align with the end of a task, an empty value with be sent instead.
These channels are logically synchronous, meaning that they have a fixed logical delay specified. Let \( \mathcal{V} \) be the set of values that can be written to a channel. Then we can define the input and output streams of a channel $q$ as follows:
\begin{equation}
q_{\text{in}}: \mathbb{N} \to \mathcal{V}, q_{\text{out}}: \mathbb{N} \to \mathcal{V}
\end{equation}

where $q_{\text{out}}(n)$ represents the value pushed onto the channel at sender clock $n$, and $q_{\text{in}}(m)$ represents the value read from the head of the channel at receive clock $m$. The logical delay \( \delta_{i \rightarrow j} \in \mathbb{N} \) relates the input and output streams of a channel from node $i$ to node $j$ as follows:

\begin{equation}
  \label{eqn:channel_delay}
q_{\text{in}}(m) = q_{\text{out}}(m - \delta_{i \rightarrow j}), \quad \text{for all } m \geq \delta_{i \rightarrow j}
\end{equation}

For \( m < \delta_{i \rightarrow j} \), the channel will contain empty data unless initial values are specified explicitly. For modules which don't have a channel dependency, $\delta_{i \rightarrow j}=\bot$

\begin{figure}[htbp]
\begin{lstlisting}[style=timetide]
module Trader:
	input price_spread : Spread;
	output order : Order;
	input fill : MatchedOrders;
	input id : integer;
	var balance : float = 1000.0 in
		var outstanding : OrderList = OrderList_create(id) in
			task(period=6, duration=3):
				if (fresh(fill)) {
					var change : float = update_orders(fill, outstanding, id) in
						balance = balance + change;
					end var;
				}
				if (fresh(price_spread)) {
					var new_order : Order = make_decision(price_spread,outstanding,balance,id) in
						if (should_do_trade(new_order)) {
							send order(new_order);
						}
					end var;
				}
			end task;
		end var;
	end var;
end module;
\end{lstlisting}
	\caption{The \acl*{TT} code of the trader module}
	\label{fig:timetide_exchange_trader}
\end{figure}

\subsubsection{Program Structure}
A \ac{TT} program consists of a network of logically synchronous threads. Different threads are distinguished by the use of the parallel operator \texttt{\small<>}. The \texttt{module}, \texttt{input}, and \texttt{output} statements are syntactic sugar for the modular instantiation of blocks of statements. \Cref{fig:hierarachy} shows the structure of a program with three distinct threads, containing the terms $w$, $u$, and $v$ respectively.
\begin{figure}[htpb]
  \begin{minipage}{0.35\linewidth}
\begin{lstlisting}[style=timetide]
{w} <> {u <> v};
\end{lstlisting}
\end{minipage}
	\caption{A hierarchy of logically synchronous threads}
	\label{fig:hierarachy}
\end{figure}

Each thread $t$ has an associated logical clock $\theta_t \in \Theta$. Threads may only communicate using named channels with a logical delay, as shared state is not possible between potentially distributed threads. Channels which are passed into a nested thread from an enclosing thread may not be used in any other thread, including the enclosing thread --- scoping of parallel threads is entirely syntactic for modularity purposes, but does not carry any special meaning from a semantic standpoint. A term within a thread is an arbitrary sequence of statements, which may include other thread declarations or a periodic task.

\subsubsection{Task Structure}
\begin{figure}[htbp]
	\begin{subfigure}[b]{0.32\linewidth}
\begin{lstlisting}[style=timetide, numbers=none]
input x : int;
output y : int;
task (duration=2,
    period=4,
    offset=1)
  send y(x + 1);
end task;
\end{lstlisting}
		\caption{Syntax Code}
	\end{subfigure}
  \hfill
	\begin{subfigure}[b]{0.32\linewidth}
\begin{lstlisting}[style=timetide, numbers=none]
loop
  sync 1;
  latch_x = x;
  sync 2;
  send y(latch_x + 1);
  sync 1;
end loop;
\end{lstlisting}
		\caption{Semantic translation}
	\end{subfigure}
	\hfill
	\begin{subfigure}[b]{0.32\linewidth}
		\centering
		\includegraphics[clip=true,trim= 0.1cm 0.0cm 0cm 0.1cm, width=\textwidth]{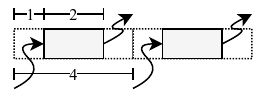}
		\caption{Input and Output sampling}
	\end{subfigure}
	\caption{A task with period 4, offset 1, and duration 2, showing input sampling and output production}
	\label{fig:tt_task}
\end{figure}

Consider the \change{task shown in \Cref{fig:tt_task}}, with period 4, offset 1, and duration 2, and its corresponding translation to semantic constructs.

The task syntax is converted semantically to a combination of various semantic constructs.
\begin{itemize}
  \item An infinite loop construct is used to describe the repeating nature of the task
  \item A \texttt{\small sync} statement is inserted at the beginning of the loop to model the offset. The sync statement progresses the logical clock by $d$, pops $d$ values from the head of each channel, and enqueues $d$ values from the data memory onto the tail of each channel
  \item A generated variable \texttt{\small latch\_x} is used to store the value of the input variable \texttt{\small x} at the sampling time of the task, according to LET semantics.
  \item A \texttt{\small sync} statement is inserted after the input sampling to progress the logical clock by the duration of the task
  \item Following the duration, the body of the task is executed in zero logical time.
  \item Finally, a \texttt{\small sync} statement is inserted at the end of the task to satisfy any extra time between the duration+offset and the period.
\end{itemize}

\change{In \Cref{fig:tt_task}, we can also see that there is no guarantee that data will be generated every logical tick.
Instead, it depends on the \emph{period} of the task.
The \texttt{\small fresh} operator is used to determine whether new data has arrived on the receiving end of a channel.}

\subsection{Pipelining}
A task may have duration longer than its period, in which case the task is pipelined. Pipelining is achieved by inserting additional parallel threads that execute the task body at the correct rate. The number of parallel threads spawned can be determined by the ratio of the \change{\ac{LCM}} of the period and duration, as shown in \Cref{eqn:threads}.
\begin{equation}
	\small
      \text{number of threads} = \frac{\text{lcm}(d,p)}{p}
      \label{eqn:threads}
\end{equation}

\begin{figure}[H]
	\begin{subfigure}[b]{0.49\linewidth}
\begin{lstlisting}[style=timetide, numbers=none]
task (duration=3, period=2, offset=0):
	t;
end task;







%padding
\end{lstlisting}
		\caption{Syntax Code}
	\end{subfigure}
  \hfill
	\begin{subfigure}[b]{0.49\linewidth}
\begin{lstlisting}[style=timetide, numbers=none]
task (duration=3, period=6, offset=0):
	t;
end task;
<>
task (duration=3, period=6, offset=2):
	t;
end task;
<>
task (duration=3, period=6, offset=4):
	t;
end task;
\end{lstlisting}
		\caption{Semantic translation}
	\end{subfigure}
	 \caption{A sugared \acl*{TT} task, and its pure translation}
	\label{fig:tt_task2}
\end{figure}

\begin{figure}[htpb]
      \includegraphics[clip=true,trim= 0.1cm 0.3cm 0cm 0.0cm, width=0.7\textwidth]{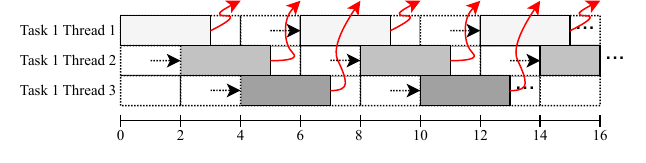}
  \caption{Timing of a task with period 2 and duration 3. Dashed input arrows to each task cycle show is multiplexed from the same input, red arrows show output production}
  \label{fig:task_timing2}
\end{figure}

The translation to parallel threads is shown in \Cref{fig:tt_task2}, and the associated timing behaviour for the task is shown in \Cref{fig:task_timing2}. A pipeline is automatically formed by the compiler such that the difference in release times of the tasks is equal to the period of the original task.

\subsection{Semantics}
\label{sec:semantics}
The formal semantics for the kernel statements are presented as program transitions using \ac{SOS} rules~\cite{plotkin2004origins} of the following form:
\begin{equation*}
      \{t, D, Q, \Theta\} \xrightarrow[]{} \{t', D', Q', \Theta'\}
\end{equation*}
where
\begin{itemize}
\item $t$ is the sequence of statements that belong to a thread,
\item $D$ is the set of values of data variables associated with $t$
  before the transition,
\item $Q$ is a global mapping from a channel label to its associated
  queue with each $q$ in $Q$ having an associated data variable
  $fresh_q$ which is true if the most recent non-empty value at the head
  of channel has not been read yet,
\item $\Theta$ is the set of clocks before the transition,
\item $\Theta'$ is the set of clocks after the transition,
\item $t'$ is the residual (remaining kernel statements) of $t$ after
  the transition,
\item $D'$ is the set of values of data variables associated with $t$
  after the transition, and
\item $Q'$ maps a channel label to its associated queue after the
  transition
\end{itemize}

Each parallel thread (e.g. $t<>...$) in \ac{TT} has an associated clock
$\theta_t$. Because most statements work within the scope of a single clock,
we use the shorthand
$\{t, D, Q\} \xrightarrow[\theta]{\theta'} \{t', D', Q'\}$ to denote that the
transition only modifies the associated clock of the thread in which $t$
is executing. Transition predicates are expressed through reduction
rules of the form
\begin{equation*}
  \frac{<pred>}{\{t, D, Q\} \xrightarrow[\theta]{\theta'} \{t', D', Q'\}}
\end{equation*}

\noindent
where the $<pred>$ must hold in order for the transition below the bar
to happen. When no such dependency exists, the bar is omitted for
simplicity. \Cref{tbl:semantics} provides an overview of the \ac{SOS}
rules for each of the kernel statements of the \ac{TT} language.

\begin{table}[htbp]
	\caption{\Acl*{SOS} Rules for \acl*{TT} Kernel statements}
      \centering
      \LARGE
      \resizebox{\linewidth}{!}{
      \begin{tabular}{|P{0.8\linewidth}P{0.8\linewidth}|}
            \hline
      {\textbf{NOTHING:}}
      \begin{equation*} \label{eq:nothing}
            \{\texttt{nothing}, D, Q \} \xrightarrow[\theta]{\theta} \{\texttt{nothing}, D, Q\}
      \end{equation*} &

      {\textbf{VAR:}}
      \begin{equation*}
            \{\texttt{x = e}, D, Q\}\xrightarrow[\theta]{\theta} \{\texttt{nothing}, D', Q\}
      \end{equation*}
      where $D' = D \cup \{x = e\}$
      \\
      {\textbf{VARDECL:}}
      \begin{equation*}
            \{\texttt{var x}, D, Q\} \xrightarrow[\theta]{\theta} \{\texttt{nothing}, D', Q\}
      \end{equation*}
      where $D' = D \cup \{x = \bot\}$
      &
      {\textbf{EXPR:}}
            \begin{equation*}
                  \{\texttt{<expr>}, D, Q\} \xrightarrow[\theta]{\theta} \{\texttt{nothing}, D, Q\}
            \end{equation*}
            \\

            {\textbf{CHANDECL:}}
            \begin{equation*}
                  \{\texttt{channel ch}, D, Q\} \xrightarrow[\theta]{\theta} \{\texttt{nothing}, D, Q'\}
            \end{equation*}
            where $Q' = Q \cup \{ch \mapsto \emptyset\}$
      &
      {\textbf{READ:}}
      \begin{equation*}
            \texttt{<ch>(x)}, D, Q \xrightarrow[\theta]{\theta} \{\texttt{<value>}, D', Q\}
      \end{equation*}
      $\text{where } D' = D \cup \{\text{fresh}_{ch} = \text{false}\}$ and $\texttt{<value>} = D[\text{buff}(x)]$

      \\
      {\textbf{SEND:}}
      \begin{equation*}
            \{\texttt{send ch(x)}, D, Q\} \xrightarrow[\theta]{\theta} \{\texttt{nothing}, D', Q\}
            \label{eq:send}
      \end{equation*}
      where $D' = D \cup \{ \texttt{buff}(ch) = x \}$
      &
      {\textbf{SYNC:}}
\begin{equation*} \label{eq:sync}
      \{\texttt{sync d}, D, Q\} \xrightarrow[\theta]{\theta'=\theta+d} \{\texttt{nothing}, D', Q'\}
\end{equation*}
where for each $i \in \{1, \ldots, d\}$:
\begin{equation*}
      \begin{aligned}
            & Freshen: D'_i = D'_{i-1} \cup \{\text{fresh}_q = \text{true} \mid q \in Q, \text{head}(q) \neq \bot\} \\
            & Pop: D'_i = D'_i \cup \{q \mapsto \text{head}(q) \mid q \in Q\} \\
            & Push: Q'_i = Q'_{i-1} \cup \{q \mapsto q \cup \text{buff}(q) \mid q \in Q\} \\
      \end{aligned}
\end{equation*}

      \\

      {\textbf{IFTRUE:}}
      \begin{equation*}
            \frac{{<}expr{>} = \texttt{true}}
            {\{\texttt{if <expr> then q else u}, D, Q\} \xrightarrow[\theta]{\theta'} \{\texttt{q}, D', Q'\}}
      \end{equation*}
      &
      {\textbf{IFFALSE:}}
      \begin{equation*}
            \frac{{<}expr{>} = \texttt{false}}
            {\{\texttt{if <expr> then q else u}, D, Q\} \xrightarrow[\theta]{\theta'} \{\texttt{u}, D', Q'\}}
      \end{equation*}
      \\
      {\textbf{LOOP:}}
      \begin{equation*} \label{eq:loop2}
            \frac{\{\texttt{\emph{t}}, D, Q'\} \xhookrightarrow[\theta]{\theta'}
                  \{\texttt{\emph{t'}}, D', Q'\}}
            {\{\texttt{loop \emph{t} end}, D, Q \}\xhookrightarrow[\theta]{\theta'}
                  \{\texttt{\emph{t'};loop \emph{t} end}, D', Q'\}}
      \end{equation*}
      where $t$ must contain a \texttt{sync} statement, or be rejected by the compiler
      &
      {\textbf{FRESH:}}
      \begin{equation*}
            \{\texttt{fresh ch}, D, Q\} \xrightarrow[\theta]{\theta} \{\texttt{<expr>}, D, Q\}
      \end{equation*}
      where $\texttt{<expr>} = D[\text{fresh}_{ch}]$
      \\


      {\textbf{SEQ1:}}
      \begin{equation*} \label{eq:seq2}
            \frac{\{\texttt{\emph{t}}, D, Q \}\xrightarrow[\theta]{\theta'}
                  \{\texttt{\emph {t'}}, D, Q'\}}
            {\{\texttt{\emph{t};\emph{u}}, D, Q \}\xrightarrow[\theta]{\theta'}
                  \{\texttt{\emph{t'};\emph{u}}, D, Q'\}}
      \end{equation*}
      &
      {\textbf{SEQ2:}}
      \begin{equation*} \label{eq:seq4}
            \frac{\{\texttt{\emph{t}}, D, Q \}\xrightarrow[\theta]{\theta}
                  \{\texttt{\emph{nothing}}, D, Q\}}
            {\{\texttt{\emph{t};\emph{u}}, D, Q \}\xrightarrow[\theta]{\theta}
                  \{\texttt{\emph{u}}, D, Q\}}
      \end{equation*}
      \\
      {\textbf{DPAR1}}:
      \begin{equation*}
            \frac{
                  \{t, D, Q\} \xrightarrow[\theta]{\theta_t'} \{t', D', Q'\} \qquad \theta_t' - \theta_u \leq \delta_{u \rightarrow t}
              }{
                  \{t, D, Q\} \mathbf{\boldsymbol{<>}} \{u, D, Q\} \xrightarrow[\{\theta_t,\theta_u\}]{\{\theta_t',\theta_u\}} \{t', D', Q'\} \mathbf{\boldsymbol{<>}} \{u, D, Q\}
              }
               \label{eq:advance}
      \end{equation*}
      &
      {\textbf{DPAR2}}:
      \begin{equation*} \label{eq:advance2}
            \frac{
                  \{t, D, Q\} \xrightarrow[\theta]{\theta_t'} \{nothing, D', Q'\}
              }{
                  \{t, D, Q\} \mathbf{\boldsymbol{<>}} \{u, D, Q\} \xrightarrow[\{\theta_t',\theta_u\}]{\{\theta_t,\theta_u\}} \{u, D, Q\}
              }
      \end{equation*}
      \\
      {\textbf{CHECKABORT:}}
      \begin{equation*}
                  \{\texttt{checkabort(c,L)}, D, Q\} \xrightarrow[\theta]{\theta} \{\texttt{nothing}, D, Q\}
      \end{equation*}
      &

      \\
      {\textbf{ABORT1:}}
      \begin{equation*}
            \frac{
                  \{t, D, Q\} \xrightarrow[\theta]{\theta'} \{t', D', Q'\} \quad
                   \texttt{!c {||} t {$\neq$} checkabort(c,L);t'}
            }{
                  \{\texttt{abort t when c:L}, D, Q\} \xrightarrow[\theta]{\theta} \{\texttt{abort t' when c:L}, D', Q'\}
            }
      \end{equation*}
      where \texttt{:L} is the unique label of this abort statement
      &
      {\textbf{ABORT2:}}
      \begin{equation*}
            \frac{
                  \{t, D, Q\} \xrightarrow[\theta]{\theta'} \{t', D', Q'\} \quad \texttt{c {\&\&} t {=} checkabort(c,L);t'}
            }{
                  \{\texttt{abort t when c:L}, D, Q\} \xrightarrow[\theta]{\theta} \{\texttt{nothing}, D', Q'\}
            }
      \end{equation*}
      \\
      \hline
	\end{tabular}}
	\label{tbl:semantics}


\end{table}


\subsubsection{Channel and Data Operations}
Every \texttt{\emph{sync}} statement inherently performs a send and a receive operation along each channel. During a send operation, the channel reads from a mailbox in the data store $D$ that contains the value to be sent at the next \texttt{\emph{sync}}. A programmer can write to this unique location using the \texttt{\small send} statement.
Rule \texttt{SEND} describes the sending of a value over a channel. The \texttt{\small send} statement is rewritten into a \texttt{\small nothing} statement after the completion of the transition. Similarly, the value of a channel is read from a unique memory location in $D$ where it was written to by the dequeue operation of the most recent \texttt{\emph{sync}} statement. For each non-empty value popped from each queue in $Q$, the value is updated in the datastore and the freshness flag is set to true. The value popped is the value written to the channel buffer exactly $\delta$ ticks ago, with respect to the sender clock.

The \texttt{\small read} statement returns the most recent non-empty value popped during a \texttt{\emph{sync}} statement, and is rewritten into its resulting value expression (Rule \texttt{READ}).

The \texttt{\small fresh} statement checks if the current value in a channel has been read previously, and returns a boolean value \texttt{\small true} if it has not been read, or \texttt{\small false} otherwise (Rule \texttt{FRESH}).
The \texttt{\small fresh} statement is only used in expressions, and is rewritten into a \texttt{\small nothing} statement.

\subsubsection{Control Flow}
Rule \texttt{SEQ1} expresses the fact that the sequence does not finish, if its
left branch, \texttt{\small\emph{t}}, does not. If the left branch advances, so does the sequence (Rule \texttt{SEQ1}). Otherwise, control will be immediately transferred to the right branch.
\texttt{\small\emph{u}}, when \texttt{\small\emph{t}} finishes (Rule \texttt{SEQ2}).

\subsubsection{Parallelism}
The clock associated with each program term is used to synchronize its behaviour with respect to others in the system. Synchronisations are required when there exists a unidirectional channel between two terms, forming a dependency. Channel delays are exact in \ac{TT}, meaning a receiving thread will not be able to progress its logical clock past $\theta_{rcv}$ until the sending term has progressed to $\theta_{snd} = \theta_{rcv}-\delta$, where $\delta$ is the channel delay (Rule \texttt{DPAR1}). If term $t$ has reduced to nothing, then only term $u$ may progress. In this case, the clock guard $\theta_t' - \theta_u \leq \delta_{u \rightarrow t}$ is not required (Rule \texttt{DPAR2}).

\subsubsection{Preemption}
\Ac{TT} provides a mechanism for preemption using the \texttt{\small abort} statement. \Ac{TT} adopts the PRET-C~\cite{andalam2009pret} approach of encoding aborts, where \texttt{\small checkabort} statements are placed into the body at compile time.
For strong aborts, a single \texttt{\small checkabort} statement immediately follows each \texttt{\small sync} statement in the term. For every weak abort, a \texttt{\small checkabort} statement is inserted directly before each \texttt{\small sync} statement in the term, except for the first. If an \emph{immediate} qualifier is added, a \texttt{\small checkabort} is inserted at the first line of the term for a strong abort, or immediately before the first \texttt{sync} for a weak abort. When aborts are nested, priority is given to the outermost abort statement through the placement of the \texttt{\small checkabort} statements. Rule \texttt{ABORT1} captures the case where the exit condition is not met, and so the term progresses. Conversely, Rule \texttt{ABORT2} captures the case where the exit condition is met and the contained term is aborted. Finally, Rule \texttt{CHECKABORT} captures the reduction of the \texttt{\small checkabort} statement itself to \texttt{\small nothing}.
These rules are not explicitly used in this work, but form the basis for future work where they are used directly for compilation and verification.

\change{
  \section{Determinism}
  \label{sec:determinism}

  Execution of a program consists of a series of \textit{reactions}, which are
  atomic units of computation in \ac{TT}, that
  delimit \emph{instantaneous} computations done between progressions of
  the logical clock of a thread, demarcated by \texttt{sync} statements,
  unless the computation terminates.

  Determinism has many interpretations, depending on the context and
  choices considered in the design of the
  system~\cite{edwards2018determinism}. \ignore{In the context of
    Logical Synchrony, we are primarily concerned with the determinism
    of individual threads as observed from the inside of the system.
    Similar to a combinational digital logic circuit, we consider a
    logical synchronous system to be deterministic even if its timing
    behavior varies from the point of view of an outside global
    observer. This is because distributed clocks may tick independently
    to some degree, as described by Rule \texttt{DPAR1}. Consequently,
    multiple global views are possible depending on the order in which
    the threads tick. We prove the determinism of logical synchronous
    systems by showing that these variations do not affect the resulting
    behaviour of the program as a whole (that is, the output will be the
    same)} Determinism in \ac{TT} is considered for each module of a
  program first. \Iac{TT} module is deterministic when starting from
  the same initial state and inputs results in the same final state
  and outputs after any number of reactions.
  
  In order to prove determinism of the overall program, we show that:
  \textcircled{1} each local module \textit{reacts} to incoming tokens
  uniquely in any logical tick, and \textcircled{2} the buffers are
  confluent irrespective of the logical execution ordering of individual
  \ac{TT} modules.
  
  \ignore{Execution of a program consists of a series of reactions,
which are atomic units of computation in \ac{TT} as \textit{reactions},
which delimit \emph{instantaneous} computations done between
progressions of the logical clock of a thread, demarcated by
\texttt{sync} statements, unless the computation terminates:}
\begin{definition}
    A \textit{reaction} consists of a sequence of instantaneous
    statements, \textit{within any branch of a parallel thread}, where
    the final transition follows a \texttt{\small sync d} statement, or
    is the term \texttt{\small nothing}. The reaction is denoted as:
    \begin{center}
      \begin{minipage}{0.95\linewidth}
        \begin{gather*}
          \{t, D, Q\} \xrightarrow[\theta]{\theta} \{t', D', Q'\}
          \xrightarrow[\theta]{\theta} \cdots \xrightarrow[\theta]{\theta} \{\texttt{\small{sync d;}}t''', D'', Q''\} \xrightarrow[\theta]{\theta+d} \{t''', D''', Q'''\} \\[1ex]
          \text{or} \quad
          \{t, D, Q\} \xrightarrow[\theta]{\theta} \{t', D', Q'\}
          \xrightarrow[\theta]{\theta} \cdots \xrightarrow[\theta]{\theta} \{\texttt{\small{nothing}}, D'', Q''\} \\[1ex]
          \text{In the following, reactions will be denoted using the following shorthand:} \\
          \{t, D, Q\} \xrsquigarrow{\theta'}{\theta} \{t', D', Q'\} \text{ where } \theta' = \theta + d \text { or } \theta' = \theta.
        \end{gather*}
      \end{minipage}
    \end{center}
\end{definition}

  \ignore{Because threads could ultimately terminate if they are not trapped within a task loop, it is necessary 
  to include in the definition above the case where the final term is \texttt{\small nothing} 
  in the definition of a reaction, as meaningful computation can still occur without reaching a \texttt{\small sync} 
  statement.}
  
  We first define the determinism of a reaction as follows.

  \begin{definition}
    \label{def:deterministic}
      Consider two reactions from the same initial state:
      \begin{equation*}
        \{t, D, Q\}\xrsquigarrow{\{\theta'\}}{\{\theta\}} \{t'_{1}, D'_{1}, Q'_{1}\}
        \textrm{\quad and \quad}
        \{t, D, Q\}\xrsquigarrow{\{\theta'\}}{\{\theta\}} \{t'_{2}, D'_{2}, Q'_{2}\},
      \end{equation*}
      Where $\{t, D, Q\}$ is the shared initial state, and
      $\{t'_{1}, D'_{1}, Q'_{1}\}$ and $\{t'_{2}, D'_{2}, Q'_{2}\}$ are
      the resulting states for two executions. Any such
      executions are deterministic \textrm{iff} the following is
      satisfied:
      \begin{equation*}
        t'_1 = t'_2, \quad D'_1 = D'_2 \text{ and }
        \operatorname{head}(q_{1}) = \operatorname{head}(q_{2})\  \forall\,
        (q_{1}, q_{2}) \in (Q'_1, Q'_2)
      \end{equation*}
      The above requires that the residual, datastore, and head of the
      incoming channel queues ($Q'_{1}$ and $Q'_{2}$) is the same, in
      spite of the timing divergence.
  \end{definition}

  We do not require the entire state of the inbound queues to be the
  same between two equivalent reactions, only their heads,
  as other values in the queue are unobservable to the receiving thread
  and hence do not affect the computation of the current reaction.
  Reactions progress in sequence to form an infinite \textit{execution}
  of \iac{TT} program:

  \begin{definition}
    An execution of a parallel \ac{TT} program starting at an initial state $\{t, D, Q\}<>\{u, D, Q\}$ is a sequence of reactions, denoted as:
    \begin{equation*}
      \{t, D, Q\}{<>}\{u, D, Q\} \xrsquigarrow{\{\theta_t',\theta_u'\}}{\{\theta\}} \{t', D', Q'\}{<>}\{u', D', Q'\}\xrsquigarrow{\{\theta_t'',\theta_u''\}}{\{\theta'\}} \{t'', D'', Q''\}{<>}\{u'', D'', Q''\}\cdots
    \end{equation*}
  \end{definition}

  Such a sequence is usually infinite, as the tasking model of \ac{TT} uses infinite loops, but could also be finite in the case of preemption. Note that despite the syntax above, a prime does not necessarily indicate a change in either parallel term, as only one of the terms may have progressed in a single reaction (unless they both reduce at the same time). 
  
  Using this definition, we define determinism for an entire \ac{TT} program as follows:

  \begin{definition}
    \label{def:deterministic2}
      \Iac{TT} \textbf{program} is deterministic iff for any two executions from the same initial state,
      \begin{enumerate}
      \item Every reaction has a corresponding reaction in the other execution with the same residual term, datastore, and head of the channel queues per \Cref{def:deterministic}.
      \item The reactions are partially ordered such that the order of each reaction associated with the same parallel thread is the same. However, the order of reactions between threads may differ.
      \end{enumerate}
  \end{definition}

  Reactions within the same thread must be ordered, as programs are naturally sequential.
  However, the causality requirement between threads is relaxed up to a channel delay, as described by the side condition in Rule \texttt{DPAR1}. Consequently, it is acceptable for two executions to have different orderings of reactions between threads, so long as causal reactions are within the channel delay of each other. 
  Despite the different orderings of reactions between threads, we prove that the behaviour of channel queues is confluent, meaning that the order in which values are pushed or popped from the queue does not affect the resulting computations.

  \begin{lemma}
    \label{lemma:confluence}
    The behaviour of each individual reaction in a single thread is the same
    for each execution regardless of the ordering of reactions between
    different threads in a parallel program, because the queues are
    confluent.
  \end{lemma}

  \begin{pf}
      Note that there are multiple rewrite choices for a parallel term:
      \begin{itemize}
      \item The left term can progress: $\{t, D, Q\} \mathbf{\boldsymbol{<>}} \{u, D, Q\} \xrightarrow[\theta]{\{\theta_t',\theta_u\}} \{t', D', Q'\} \mathbf{\boldsymbol{<>}} \{u, D, Q'\}$
      \item The right term can progress: $\{t, D, Q\} \mathbf{\boldsymbol{<>}} \{u, D, Q\} \xrightarrow[\theta]{\{\theta_t,\theta_u'\}} \{t, D, Q'\} \mathbf{\boldsymbol{<>}} \{u', D', Q'\}$
      \item Both terms progress: $\{t, D, Q\} \mathbf{\boldsymbol{<>}} \{u, D, Q\} \xrightarrow[\theta]{\{\theta_t',\theta_u'\}} \{t', D', Q'\} \mathbf{\boldsymbol{<>}} \{u', D', Q'\}$
      \end{itemize}

      The value of the channel at the end of a logical tick may differ as a consequence of which arm is reduced.
      However, behaviour of any one reaction only depends on the values popped from the heads of the queues. Values may only be inserted in-order by appending to the tail of the queue (Rule \texttt{SYNC}).
      The only case where the tick of $u$ could affect $t$ is if the queue is empty, in which case the transition of $t$ would not be enabled in the first place. Thus, when $t$ eventually ticks, the value it pops from each $head(q)$ is always the same regardless of the different possible transitions.
  \end{pf}

  Although the relative ordering of reactions is confluent, the individual reactions themselves must also be deterministic per \Cref{def:deterministic}. We prove this by showing that each individual rewrite within a reaction is unique:

  \begin{lemma}
    \label{lemma:singlevalue}
      A reaction in \iac{TT} program is finite and single-valued, meaning that it always reduces to a unique residual term
  \end{lemma}
  \begin{pf}
      By structural induction on the term \( t \).

      \begin{itemize}
      \item \textbf{Base case:}
        Suppose \( t = \texttt{\small nothing} \) or \( t = \texttt{\small sync d;t'} \).
        In both cases, the reaction terminates immediately yielding \texttt{\small nothing} or the unchanged term \( t' \) respectively.

      \item \textbf{Case \( t = \texttt{\small loop t' end}\):}
        By Rule \texttt{LOOP}, all loops must contain at least one \texttt{\small sync} in every iteration. Therefore, any sequence of reductions from \( t \) will eventually encounter a \texttt{\small sync} statement, ending the reaction for the current tick.

      \item \textbf{Case \( t = u <> v \):}
        All possible interleavings of sub-terms are confluent per \Cref{lemma:confluence}, and each sub-term \( u \) and \( v \) will eventually reduce to a unique residual term.

      \item \textbf{Other Cases:}
        Each base statement has a unique rewrite rule, except for branching statements \texttt{\small abort} and \texttt{\small if}, which may only vary as a result of input channel data which is confluent per \Cref{lemma:confluence}. Hence, the residual is still uniquely determined. \qedhere
      \end{itemize}
  \end{pf}

  Thus, if each individual reaction is deterministic, and communications between reactions are confluent, then it follows that the overall global behaviour of the program is deterministic.

  \begin{thm}
    A program written with the \ac{TT} semantics is deterministic.
  \end{thm}
  \begin{pf}
    By contradiction. Assume that a non-deterministic program exists. This
    implies that in at least one reaction in the execution, a transition
    from a known state leads to two different subsequent states, i.e.:

    \begin{equation*}
      \{t, D, Q\} \xrsquigarrow{\{\theta'\}}{\{\theta\}} \{t_1, D_1, Q_1\}
      \quad \text{and} \quad
      \{t, D, Q\} \xrsquigarrow{\{\theta'\}}{\{\theta\}} \{t_2, D_2, Q_2\}
    \end{equation*}

    Per \Cref{def:deterministic}, this means that the residual terms
    \( t_1 \) and \( t_2 \) must be different, or the data store \( D_1 \)
    and \( D_2 \) must differ, or the heads of the channel queues must
    differ i.e.
    $\forall\, (q_1, q_2) \in (Q'_1, Q'_2), \operatorname{head}(q_1) =
    \operatorname{head}(q_2)$

    However, each reaction has a unique rewrite as shown in
    \Cref{lemma:singlevalue}, which is also the only way the data store is
    updated. Furthermore, the heads of channel queues are always
    confluent per \Cref{lemma:confluence}. Thus, the residual terms, data
    store, and heads of channel queues must be the same in both cases,
    contradicting the assumption that the program is non-deterministic.
  \end{pf}

  This definition of determinism has one major catch: the effects of physical time are disregarded. This is a recurring theme in the Logical Synchrony model, where we detach our synchronisation between threads from the physical time of the system. Interactions with the physical world are non-trivial and will be the subject of future work. However, we posit that there are still many applications where physical time is less critical than logical correctness, such as in distributed computation.
}


\section{Synchronous Execution and Verification}
\label{sec:verification}

\acl{TT} models a purely synchronous abstraction of a distributed system. We have implemented a source-to-source compiler to convert \acl{TT} into equivalent Esterel code. This supports two modes: a centralised mode which produces a single program for verification, and a distributed model which produces multiple communicating programs for deployment.

\subsection{Distributed Target}
In distributed mode, a separate Esterel module is produced for each thread of the \acl{TT} program, which is subsequently compiled into a library that exposes a logical tick function.
\change{
	To execute these threads/modules $\Tau$ over a distributed \ac{LSN} $\mathcal{L}$, we need to define a mapping function $\Gamma : \Tau \rightarrow V$ which says which \ac{LSN} node (in $V$) is used to execute each thread (in $\Tau$).
	This mapping could be either manually defined or automatically generated (using techniques such as \ac{ILP}~\cite{floudas2005mixed} or \ac{SA}~\cite{attiya2006task}) based on some constraints.
	For all communication channels, the delay through the \ac{LSN} must be less than or equal to the \ac{TT} delay, i.e. $\forall_{\tau_1, \tau_2 \in \Tau} : \lambda_{\Gamma(\tau_1) \rightarrow \Gamma(\tau_2)} \leq \delta_{\tau_1 \rightarrow \tau_2}$.
	In the worst case, threads may need to be scheduled on the same node which incurs zero communication delay (but reduces parallelism).
	Additional constraints could be added based on system requirements, such as requiring tasks be executed on specific nodes due to available resources.
}

\change{As a proof of concept in this work, we developed a} lightweight wrapper \change{that} handles channel communication over network sockets. Each channel is modelled as a token-pushing FIFO buffer situated on the receiver end. To form the logical delays, each FIFO is pre-populated with a number of initial tokens. The main function simply performs a loop which pops from each inbound buffer, invokes the tick, and writes to each output socket. An abridged version of the generated wrapper is shown in \Cref{lst:generated_c}.

\begin{figure}[htbp]
	\begin{lstlisting}[style=c]
		init_price_spread("tcp://localhost:1234"); // initialise channel
		while(1) {
			Signal price_spread = get_price_spread();	// Blocking read
			if(price_spread.is_present()) { // Value or empty frame
				in_price_spread(price_spread.val());
			}
			auto module_outputs = module_run(); // Run the esterel module
			send_order(module_outputs.order);// Write to outbound channels
		}
	\end{lstlisting}
	\caption{Abridged generated C++ wrapper}
	\label{lst:generated_c}
\end{figure}

\begin{lemma}
    The use of a \ac{FIFO} with initial tokens faithfully implements a logical delay as defined in the \ac{TT} semantics (Rules \texttt{DPAR1} and \texttt{DPAR2}).
    \label{lemma:initial_token}
\end{lemma}

\begin{pf}
	By induction on the receiver clock $\theta$.
\end{pf}

\subsection{Centralised Target}
In the centralised compilation from \ac{TT}, channels are synthesized as a number of intermediate signals, which form a shift register as shown in \Cref{fig:timetide_to_esterel}

\begin{figure}[htbp]
	\begin{subfigure}[b]{0.49\linewidth}
\begin{lstlisting}[style=timetide]
chan <id> delay <n> in
	run <module>[<id>/<id>];
	<>
	run <module>[<id>/<id>];
end chan







%padding
\end{lstlisting}
		\caption{\Acl*{TT}}
		\label{fig:timetide_to_esterel:timetide}
	\end{subfigure}
	\begin{subfigure}[b]{0.49\linewidth}
\begin{lstlisting}[style=esterel,numbers=right]
signal <id>_0,...<id>_<n> in
	run <module>[signal <id>_0];
	||
	run <module>[signal <id>_n];
	||
	loop
		emit <id>_<n>(pre(<id>_<n-1>));
		pause;
	end;
	||
	loop
		emit <id>_<n-1>(pre(<id>_<n-2>))
	...
\end{lstlisting}
		\caption{Esterel}
		\label{fig:timetide_to_esterel:esterel}
	\end{subfigure}
    \caption{Translation of \acl*{TT} channels to Esterel signals}
    \label{fig:timetide_to_esterel}
\end{figure}

\begin{lemma}
  The use of cascaded \texttt{pre} operators in the centralised Esterel code faithfully implements a logical delay as defined in the \acl{TT} semantics.
  \label{lemma:pre}
\end{lemma}
\begin{pf}
	By induction on the number of cascaded \texttt{pre} operators $k$.
\end{pf}

\subsection{LSN-Compatible Architectures}
\label{sec:mapping}
The deterministic behaviour of \iac{TT} program is only relevant if it is executed \change{on} a suitable \change{synchronisation layer}.
To this end, \change{we define \emph{\ac{LSN}-compatible architectures}} which allow for such deterministic execution. The minimum requirement for \iac{LSN}-compatible architecture is that the tick function from every thread in the \ac{TT} specification may be called for an $n$th time only if the values from every inbound channel that were sent at logical time $n{-}\delta$ are available, where $\delta$ is the logical delay of that channel. It follows that because \iac{TT} program's behaviour solely varies upon the inputs it receives, then any architecture which guarantees the same input sequence will behave the same.

A number of approaches satisfy this requirement\change{, and it is left up to the specific implementation to decide how it handles the buffers}. For example, the bittide approach~\cite{lall2022modeling} is completely blocking-free \change{(push-model)}, since it manages potential buffer overflow by self-balancing the clock speed of each distributed node. In comparison, the Kahn-like \acp{FFP}~\cite{kenwright2024logical} manages buffer overflow by using both blocking reads and writes \change{(sometimes acting as a pull-model)} across channels.
\change{Such blocking-based implementations can allow for \iac{LSN} to be implemented over generic networks such as TCP/IP.}

\change{
	\subsection{Relation to the Physical World}

	\Ac{TT} programs live in the world of logical synchrony, meaning that the only reference to time is that of each node's local (synchronous) clock.
	While this is an advantage in the distribution, determinism, and verification of these programs, it typically restricts them to \emph{closed} systems with no interface to the physical world.
	For some use-cases this is acceptable, such as closed-loop simulations, however for others this restriction needs to be relaxed.

	There is one important question to keep in mind while
	contemplating the physical world: how does one guarantee determinism when each node has its own different view of the physical world, even at the same logic time?
	Allowing each individual node to interact with the physical
	world around it would lead to non-determinism, as signals which occur at the same physical time may be seen at different logical times, or vice versa.
	While it may be possible to bound this discrepancy using the \ac{LSN} communication delays and the \ac{WCRT} of the application, this still poses problems.

	To address this, we allow only a single node to interact with the environment, for physical time, inputs, or outputs.
	As a result, there is only one ``view'' of the physical world for the \ac{TT} program, removing any potential paradoxical cases.
	For any other node which wishes to interact with the outside
	world, signals must travel from the node with the outside
	view, via paths of channels with defined logical delays in order to maintain determinism.
}

\subsection{Model Checking of \acl{TT} Programs}
Because \acl{TT} programs are semantically synchronous, despite the distributed execution, they may be easily verified using existing approaches as if they were a single program, due their equivalence as shown in \Cref{thm:timetide_equivalence}. We do not consider the possibility of implementation-specific interference such as resource contention or pre-emption, as these are fundamentally implementation details to be solved separately of the \acl{TT} model itself.

\begin{thm}
	The execution of the distributed target for \iac{TT} program is equivalent to that of the centralised target.
	\label{thm:timetide_equivalence}
\end{thm}
\begin{pf}
	By induction on the logical clock $\theta$ of a thread, and given \Cref{lemma:initial_token,lemma:pre}:
	\begin{itemize}
		\item At $\theta=0$ the initial states of both implementations are identical, and initial inputs match. The task code is identical in both implementations, thus outputs are identical.
		\item Assume that for $\theta=k$ the internal state of both implementations are identical. At $\theta=k+1$, the output is computed solely from the current state and the inputs of each inbound channel. Given that both implementations are equivalent, the inputs to the tasks are identical, hence so are the outputs. \qedhere
	\end{itemize}
	\label{thm:equivalence}
\end{pf}

\subsubsection{Modelling}
Historically, safety properties of synchronous programs are verified using \emph{synchronous observers}, which are programs that are composed in parallel with the main program that raise an alarm upon property violation. As opposed to an equally expressive \ac{LTL} property, the condition of a synchronous observer is written in the same language as the program, making them simple to write and understand.

Consider again the trading system from \Cref{sec:app_examples}. A safety property for this system may be that if a trader submits an order, it will not be missed by the centre.
Intuitively, we might think this could be a problem since the trader runs more frequently than the exchange. However, this is addressed through the \texttt{fresh} operator.
This property is encoded using synchronous observer shown in \Cref{lst:observer}. One way of encoding this is to ensure that the trade ID of the last order received by the exchange from a trader is exactly one greater than the previous order, otherwise it is implied that an order was missed. We compose the observer module in distributed parallel with the main program, as shown in \Cref{fig:observer_compose}. Subsequently, in a model checker we simply verify that \texttt{PROPERTY\_VIOLATED} is never emitted.

\begin{figure}[htbp]
\begin{lstlisting}[style=timetide]
module observer:
  input ack : integer;
  output PROPERTY_VIOLATED : boolean;
  var last_order := -1 : integer in
	  task(period=1, duration=1, offset=0):
	    if (fresh(ack)) {
	      if (ack != last_order+1) {
	        send PROPERTY_VIOLATED(true);
	      }
	      last_order := ack;
	    }
	  end task;
  end var;
end module
\end{lstlisting}
	\caption{Safety Property Observer}
	\label{lst:observer}
\end{figure}
\begin{figure}[htbp]
\begin{lstlisting}[style=timetide, linewidth=0.99\textwidth]
...
channel ack : integer delay 1;
channel PROPERTY_VIOLATED : boolean delay 0;
run Center(t1_order, t1_trade, t2_order, t2_trade, t1_spread, t2_spread, ack);
<>
run Trader(t1_spread/price_spread, t1_order/order, t1_trade/trade, 0/id);
<>
run Trader(t2_spread/price_spread, t2_order/order, t2_trade/trade, 1/id);
<>
run observer(ack, PROPERTY_VIOLATED);
\end{lstlisting}
	\caption{A synchronous observer composed with the main program}
	\label{fig:observer_compose}
\end{figure}

These properties which pertain to the timing of events are much easier to verify than more general safety properties. Because the \ac{LET} tasks execute periodically, they necessarily form a repeated hyperperiod of execution~\cite{gunzel2021timing}, following some initial prelude. Thus, it is sufficient to verify the program over a single (non-prelude) hyperperiod, which captures all possible interactions between the tasks.

\subsubsection{Verification Results}
For demonstration, we have implemented several synchronous observers for applications including the trading system as well as a model of basic cruise controller, shown in \Cref{fig:timetide_cruise:top}. For this work we simply use the bounded model checker CBMC~\cite{kroening2014cbmc} on the actual generated code, which is sufficient for safety and bounded-liveness properties. An example of the bounded liveness property is shown in \Cref{fig:timetide_cruise:liveness}.
\begin{figure}[htpb]
	\centering
	\begin{subfigure}[b]{0.45\linewidth}
\begin{lstlisting}[style=timetide]
chan speed : float delay 1,
  thtl     : float delay 2,
  rpm      : float delay 1,
  setpoint : float delay 1,
in
  run ShaftSensor(speed, rpm);
  <>
  run Ctrl(speed, thtl, setpoint);
  <>
  run Motor(thtl, rpm);
end chan;



%padding
\end{lstlisting}
	\caption{The toplevel code of the cruise control}
	\label{fig:timetide_cruise:top}
\end{subfigure}
\begin{subfigure}[b]{0.54\linewidth}
\begin{lstlisting}[style=timetide, numbers=right]
module LivenessObserver:
	input setpoint, speed : float;
	output LIVE_VIOLATED : boolean = false;
	var counter := 0 : integer in
		task(period=1,duration=1):
			if (fresh(setpoint)) { counter := 0; }
			if (setpoint != current_speed) {
				counter = counter + 1;
			} else { counter = 0; }
			if (counter >= 10) {
				send LIVE_VIOLATED(true);
			}
		end task
	end var
end module
    \end{lstlisting}
	\caption{The \acl*{TT} code of the bounded liveness property}
	\label{fig:timetide_cruise:liveness}
\end{subfigure}
\caption{The cruise control example and a synchronous observer}
\label{fig:timetide_cruise}
\end{figure}

Other property implementation details are omitted but are available in the \ac{TT} repository\footnote{\emph{URL removed for blind review}}. Results are shown in \Cref{tab:verification_results}, including a property which fails to validate the verification process. Because this implementation is a simple proportional controller, there is no explicit prevention of negative speed, which was correctly identified by the model checker.

\begin{table} [htbp]
\centering
\footnotesize
\caption{Verification Results}
\begin{tabular}{|l|c|c|c|}
\hline
\textbf{Program} & \textbf{Property Description} & \textbf{Result} \\
\hline
\texttt{Trading} & No missed orders & PASS \\\hline
\texttt{Trading} & Stock cannot be overtraded & PASS \\\hline
\texttt{Cruise Controller} & Not more than 5\% over target speed & PASS  \\\hline
\texttt{Cruise Controller} & Target speed reached within 10 ticks & PASS \\\hline
\texttt{Cruise Controller} & Speed can never be negative & FAIL \\
\hline
\end{tabular}
\label{tab:verification_results}
\end{table}

\section{Results}
\label{sec:comparison}
As the closest related language, we compare the performance and correctness of \ac{TT} with that of \ac{LF}. The example applications were implemented in both languages, using the same host calls, and the throughput and trace equivalence of the two languages were compared. All benchmarks were run locally on a $2021$ Macbook Pro with a $10$-Core M1 Pro chip. Network latency was simulated using the tool \emph{Comcast}~\cite{comcast_tool}.  
To create an equivalent \ac{LF} system, we have the following translation rules: Each distributed thread becomes a reactor with the same inputs, outputs, and variables. A `fresh' boolean variable is added for each input. Each task becomes a periodic reaction, with an additional reaction to freshen each input. Each logical tick is translated into 1\,ms of physical time. The important characteristic is the relative task lengths, so the unit of time is not important.
A minimal example of the translation from \ac{TT} to \ac{LF} is shown in \Cref{fig:timetide_to_lf}.
\begin{figure}[htbp]
	\begin{subfigure}[b]{0.42\linewidth}
\begin{lstlisting}[style=timetide]
module example:
	input a : integer;
	output b : integer;
	var x : integer = 0 in 
		task(duration = 2, period = 3):
			if fresh(a){
				x = a;
				send b(x);
			}
		end task;
	end var;
end module


%padding
\end{lstlisting}
		\caption{\Acl*{TT}}
		\label{fig:timetide_to_lf:timetide}
	\end{subfigure}
	\begin{subfigure}[b]{0.45\linewidth}
\begin{lstlisting}[style=lf,numbers=right]
reactor example{
	input a : int;
	output b : int;
	state x : int = 0;
	state fresh_a : bool = false;
	timer t(0,3ms);
	reaction(t) a -> b {=
		if (fresh_a) {
			self->x = a;
			fresh_a = false;
			lf_set(b, self->x)
		}
	=};
	reaction(a) {=self->fresh_a=true;=};
}
\end{lstlisting}
		\caption{\Acl*{LF}}
		\label{fig:timetide_to_lf:lf}
	\end{subfigure}
	\caption{Translation of \iacl*{TT} module to \acl*{LF}}
	\label{fig:timetide_to_lf}
\end{figure}

Furthermore, instantiations in \ac{TT} are directly translated to reactor instantiations in \iac{LF}, as shown in \Cref{fig:timetide_to_lf2}. The task duration is added to the transmission delay in the \ac{LF} program, as reactors take zero logical time to execute.

\begin{figure}[htbp]
	\begin{subfigure}[b]{0.49\linewidth}
\begin{lstlisting}[style=timetide]
channel ch1 : integer delay 3;
run example[ch1/b] <> run example[ch1/a];
%padding
\end{lstlisting}
		\caption{\Acl*{TT}}
		\label{fig:timetide_to_lf2:timetide}
	\end{subfigure}
	\begin{subfigure}[b]{0.49\linewidth}
\begin{lstlisting}[style=lf,numbers=right]
example e1 = new example();
example e2 = new example();
e1.b -> e2.a after 5ms; % dur. + delay
\end{lstlisting}
		\caption{\Acl*{LF}}
		\label{fig:timetide_to_lf2:lf}
	\end{subfigure}
	\caption{Translation of module instantiation from \acl*{TT} to \acl*{LF}}
	\label{fig:timetide_to_lf2}
\end{figure}

Using our financial trading example, we ensure that the resulting output trace between the two languages is equivalent. The external function calls are exactly the same, so the only difference is the coordination infrastructure provided by the language.
When applying the same seed to the random number generators in each language and logging the outputs, we confirm that the observable traces of the two implementations are identical.

\Ac{LF} has two distributed coordination schemes: the default, which coordinates distributed threads with a central arbiter, and a decentralised peer-to-peer coordinator. To avoid confusion with the non-distributed variant, we will refer to these two flavours as \change{\textit{Arbiter~\acs*{LF}}} and \change{\textit{P2P~\acs*{LF}}}. Both of these are compared to the \ac{TT} approach. The \ac{TT} approach is more similar to the \change{P2P~\acs*{LF}} approach. However, \change{P2P~\acs*{LF}} cannot run unrestricted by physical time, unlike Arbiter mode, as the P2P distribution relies on physical timing to synchronise. Consequently, to give a fair benchmark we iteratively increased the physical-time gearing \change{(slowing down the program)} of the \change{P2P~\acs*{LF}} nodes until \change{until the \ac{LF} program stopped reporting timing errors}, in order to get tight bounds on the fastest execution rate.  We ran the financial trading system for $10,000$ logical ticks using both the \ac{TT} and \ac{LF} generated code and measured the total execution time. The results are shown in \Cref{fig:throughput}, averaged over 3 executions per delay variation.

\begin{figure}[htbp]
	\centering
	\small
	\begin{tikzpicture}
		\begin{axis}[
			legend style={font=\footnotesize},
			xlabel={Simulated Delay (\unit{\milli\second})},
			ylabel={Execution Time (\unit{\second})},
			legend pos=north west,
			grid=major,
			width=0.65\columnwidth, 
			height=0.35\columnwidth, 
		]
			\addplot[color=blue, mark=square*] coordinates {
				(0, 4.186)
				(1, 5.805)
				(2, 7.621)
				(3, 9.157)
				(4, 11.078)
				(5, 13.151)
				(6, 14.248)
				(7, 14.828)
				(8, 15.917)
				(9, 18.186)
				(10, 20.384)
				(11, 22.010)
				(12, 24.074)
			};
			\addlegendentry{\Acl*{TT}}
			
			\addplot[color=red, mark=triangle*] coordinates {
				(0, 15.836)
				(1, 20.152)
				(2, 23.797)
				(3, 28.196)
				(4, 31.958)
				(5, 35.844)
				(6, 39.607)
				(7, 42.936)
				(8, 47.226)
				(9, 51.694)
				(10, 56.316)
				(11, 60.765)
				(12, 65.150)
			};
			\addlegendentry{Arbiter~\acs*{LF}}

			\addplot[color=purple, mark=o] coordinates {
				(0, 20.000)
				(1, 25.000)
				(2, 25.000)
				(3, 27.500)
				(4, 27.500)
				(5, 30.000)
				(6, 30.000)
				(7, 32.500)
				(8, 32.500)
				(9, 35.000)
				(10, 35.000)
				(11, 37.500)
				(12, 40.000)
			};
			\addlegendentry{P2P~\acs*{LF}}
    	\end{axis}
	\end{tikzpicture}
	\caption{Throughput comparison between \acl*{TT} and \acl*{LF}}
	\label{fig:throughput}
\end{figure}

Although \ac{LF} is undoubtedly the more mature compiler and execution time appears to be roughly linear for both tools, the throughput of the \ac{TT} generated code is significantly (around $3\times$) faster than the Arbiter variant and almost twice as fast as the P2P variant. This is likely due to two major factors:
\begin{enumerate}
	\item \Ac{TT} compiles to baremetal C and communicates over a lightweight wrapper. In comparison, \ac{LF} relies on a runtime which relates logical and physical time. Despite disabling physical time restrictions, the overhead of the runtime remains.
	\item \Ac{TT} automatically pipelines its execution when transmission delays are specified, which the arbiter flavour of \ac{LF} doesn't do. 
\end{enumerate}

\begin{figure}[htpb]
\includegraphics[clip=true,trim=0cm 0.05cm 0cm 0cm,width=0.7\linewidth]{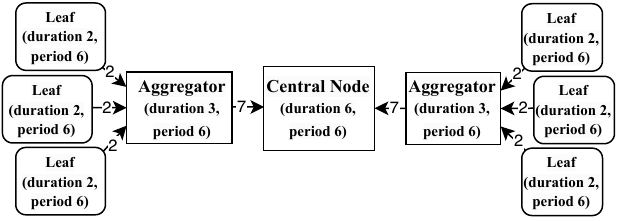}
\caption{\change{Block diagram of a sensor network with two aggregator nodes which each have three leaf nodes}}
\label{fig:sensor_network}
\end{figure}

\change{
To examine \ac{TT}'s capability in building large-scale systems, we have implemented a model of a sensor network.
This sensor networks consists of \textit{leaf nodes} which sense data from the environment, \textit{aggregator nodes} which combine data from local leaf nodes into a single data point, and a \textit{central node} which ultimately receives data from all aggregator nodes. An example with two aggregator nodes, each with three leaf nodes, is shown in \Cref{fig:sensor_network}. Compared with the trading example, this example demonstrates hierarchy in \ac{TT}. Moreover, the use of sensor nodes with a low latency to a local compute node, and a larger latency link to a central node, is a common pattern in modern edge computing.
Here, we match the periods of all nodes to the same value, so that the production and consumption of sensor data is matched. An abridged version of the \ac{TT} code is shown in \Cref{lst:sensor_network}.
}

\begin{figure}[htbp]
\begin{subfigure}[b]{0.46\linewidth}
\begin{lstlisting}[style=timetide]
module toplevel:
	const AGGS : int = 2;
	channel from_agg : int[AGGS] delay 7;

	run CentralNode[from_agg] 
	<>
	pareach i in AGGS {
		run Aggregator[i, from_agg[i]] 
	}
end module;




%padding
\end{lstlisting}
		\caption{\change{Top level module}}
	\end{subfigure}
	\begin{subfigure}[b]{0.52\linewidth}
\begin{lstlisting}[style=timetide,numbers=right]
module Aggregator:
	const LEAVES : int = 3;
	output from_agg : int;
	channel from_leaf : int[LEAVES] delay 2;

	pareach i in LEAVES {
		run LeafNode[i, from_leaf[i]] 
	} <>
	task(duration=3, period=6):
		var aggregate : float = aggregate_readings(from_leaf) in
				send from_agg(aggregate);
		end var;
	end task;
end module;
\end{lstlisting}
		\caption{\change{Aggregator Node}}

\end{subfigure}
\caption{\change{The sensor network example in \acl*{TT}}}
\label{lst:sensor_network}
\end{figure}

%
%

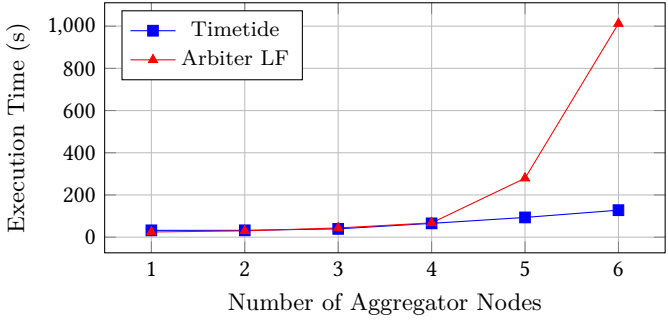
\begin{figure}[htbp]
	\centering
	\small
	\begin{tikzpicture}
		\begin{axis}[
			legend style={font=\footnotesize},
			xlabel={Number of Aggregator Nodes},
			ylabel={Execution Time (\unit{\second})},
			legend pos=north west,
			grid=major,
			width=0.65\columnwidth, 
			height=0.35\columnwidth, 
			ytick={0,200,400,600,800,1000},
		]
			\addplot[color=blue, mark=square*] coordinates {
				(1, 32.3) 
				(2, 32.4) 
				(3, 39.38) 
				(4, 65.43) 
				(5, 93.53) 
				(6, 127.91) 
			};
			\addlegendentry{\Acl*{TT}}

			\addplot[color=red, mark=triangle*] coordinates {
				(1, 24.6)
				(2, 29.9)
				(3, 43.7)
				(4, 67.73)
				(5, 279.319)
				(6, 1012.11)
			};
			\addlegendentry{Arbiter~\acs*{LF}}
			
    	\end{axis}
	\end{tikzpicture}
	\caption{\change{Sensor Network Throughput on a 10-Core M1 Pro Macbook with Varying Aggregator Nodes}}
	\label{fig:throughput3}
\end{figure}

\change{

To demonstrate how \ac{TT} handles programs with a larger number of tasks, we measure the execution time while varying the number of aggregator nodes (\texttt{AGGS}), where each additional aggregator node adds four tasks to the system.
The results are shown in \Cref{fig:throughput3}, using a latency of \qty{500}{\milli\second} between each aggregator node and the main hub.
}

\change{
For one and two aggregator nodes we have 5 and 9 threads respectively, meaning that each thread can be assigned to a core on the CPU.
Consequently, we see no performance penalty, even with the additional communication.
For three or more aggregator nodes, the number of threads exceeds the number of cores and consequently we see a relatively linear increase in execution time, showing that system performance does not degrade exponentially with additional threads.
\Ac{LF} will spawn local worker threads for instantiated reactors within other reactors, but not in a distributed fashion, and seems to scale poorly with the number of communicating threads.
For comparison, running the same 6-arbiter program on the \emph{centralized} \ac{LF} runtime yielded an execution time of just \qty{66.6}{\second}.
}

We also compare the two toolchains by their compilation speed, number of lines of code (albeit, according to our own translation rules), and the size of the generated binaries. Binary size is the sum of all produced binaries for a distributed program, however the separately-installed \acf{LF} Runtime Interface is not included. For these comparisons, we use our financial exchange \change{and sensor network examples} along with two (periodic) examples from the \ac{LF} playground. The results are shown in \Cref{table:comparison}.

\begin{table}[htbp]
    \caption{Comparison of Compilation Speed, Number of Lines of Code, and Binary Size}
	\scriptsize
    \centering
    \begin{tabular}{lcccccccc}
        \toprule
        & \multicolumn{2}{c}{Exchange} & \multicolumn{2}{c}{CAL} & \multicolumn{2}{c}{Rosace} & \multicolumn{2}{c}{\change{Sensor Network}} \\
        \cmidrule(r){2-3} \cmidrule(r){4-5} \cmidrule(r){6-7} \cmidrule(r){8-9}
        & {TT} & {LF} & {TT} & {LF} & {TT} & {LF} & \change{TT} & \change{LF}\\
        \midrule
        Num. Lines & {\textasciitilde}110 & {\textasciitilde}130 & {\textasciitilde}50 & {\textasciitilde}50 & {\textasciitilde}400 & {\textasciitilde}400 & \change{{\textasciitilde}100} & \change{{\textasciitilde}100}\\
		Num. Threads & 3 & 3 & 4 & 4 & 11 & 4 & \change{9} & \change{3}\\
        Sum Bin. Size (\unit{\kilo\byte}) & 1938 & 337 & 2045 & 396 & 6250 & 425 & \change{4780} & \change{298} \\
        Compile Time (\unit{\second}) & 5.4 & 22.3 & 6.0 & 22.8 & 26.4 & 26.8 & \change{17.1} & \change{21.5} \\
        \bottomrule
    \end{tabular}
    \label{table:comparison}
\end{table}

The compilation speed of \ac{TT} is significantly faster than \ac{LF} for these examples. However, the \ac{LF} compilation time does not tend to vary significantly for different programs --- much of the time is taken compiling the appreciable runtime. In comparison, the syntax translation from \ac{TT} to Esterel takes negligible time and consequent Esterel compilation is similarly rapid. The number of lines of code is not meaningfully different, as the constructs of each language have similarities. The size of the generated binaries is somewhat larger in \ac{TT} than in \ac{LF}, \change{likely because the code generated by the dated Esterel v5 compiler is not particularly compact}. Interestingly, the \ac{TT} implementations of the Rosace \change{and Sensor Networks} benchmarks produce many more distributed threads, \change{due to the nesting of instantiations actually spawning additional threads}. In comparison, only top-level instantiations are distributed in \iac{LF} system. In any case, no concerning behaviour was observed in the above metrics for either language; the primary difference is the logical versus physical model of synchronisation. 

Ultimately, we are comparing the performance of \ac{TT} within its niche to \ac{LF} operating outside its intended domain of real-time systems. However, we are not aware of any other languages which provide deterministic execution across a distributed system to compare against. Many of the \ac{LF} playground examples can simply not be expressed in \ac{TT} due to the lack of support for input from the environment. Hence, the \ac{TT} model is less expressive than \ac{LF}, in exchange for its simplified synchronisation model. Thus, we conclude that the \ac{TT} model of logical synchrony is the most performant choice for systems which require determinism and high throughput, but do not necessarily have real-time or reactivity requirements.

\section{Related Work}
The synchronous paradigm was independently introduced through three foundational languages: Esterel~\cite{berry1992esterel},
Lustre~\cite{halbwachs1991synchronous} and Signal~\cite{benveniste1991synchronous}.
These rely on the \emph{synchrony hypothesis}, assuming a reactive system operates infinitely fast relative to its environment. A logical tick function is repeatedly invoked, which samples inputs from the environment and computes the outputs. Synchronous programs typically \emph{compile away concurrency} to produce sequential code which scales well, even for large systems. However, distribution of synchronous programs is challenging~\cite{baudisch2010dependency,girault2005survey}, due to the expense of distributing a global clock. Many works instead aim to focus on making synchronous programs \textit{latency-insensitive}~\cite{carloni2001theory} to avoid the need for synchronisation, however this is limited to a small subset of synchronous programs. The \emph{N-synchronous model}~\cite{cohen2006n,mandel2010lucy} relaxes synchrony by allowing threads to be desynchronised by at most \texttt{n}-ticks using a FIFO buffer. In N-synchrony, buffer sizes
need to be computed based on a schedule and synthesized instead being specified, unlike \acf{TT}. Our approach relies on implicit buffers, which are left to the communication protocol of the underlying implementation, such as bittide~\cite{lall2022modeling}
or Finite FIFO Platforms (FFP)~\cite{tripakis2008implementing}.
Table~\ref{table:deterministic-langaues} summarises some languages that enable deterministic distribution, based on how they model time, synchronise, specify task duration, and express communication delay. 

\begin{table}[htpb]
      \caption{Deterministic languages for distributed systems}
      \centering
      \footnotesize
      \begin{tabularx}{\textwidth}{|>{\raggedright\arraybackslash}p{0.2\textwidth}|>{\raggedright\arraybackslash}X|>{\raggedright\arraybackslash}X|>{\raggedright\arraybackslash}X|>{\raggedright\arraybackslash}p{0.18\textwidth}|}
            \hline
            {\bf Language}                                & {\bf Model of Time} & {\bf Synchronisation} & {\bf Task duration} & {\bf Latency} \\
            \hline
            Multiclock-Esterel~\cite{berry2001multiclock} & logical             & HW sync.       & multiples of ticks  & implicit      \\
            \hline
            Giotto~\cite{henzinger2001giotto}             & physical             & clock sync.           & zero               & instantaneous \\
            \hline
            PsyC~\cite{siron2023formal}                   & logical             & clock sync.           & multiples of ticks               & instantaneous \\
            \hline
            Lingua Franca~\cite{lohstroh2021toward}       & logical+physical    & clock sync.           & zero               & optional fixed delay \\
            \hline
            Timetide                                     & logical             & logical sync.     & multiples of ticks               & fixed delay   \\
            \hline
      \end{tabularx}
      \label{table:deterministic-langaues}
\end{table}

The languages Giotto and PsyC are both used for the modelling of \ac{LET} systems, but assume insignificant communication delays for distribution, which is not realistic but is instead abstracted away during scheduling on a physical device. The exception to this is the recent federated flavour of Lingua Franca~\cite{lohstroh2024deterministic}, which specifies transmission delays but either requires a centralised coordinator or leverages strong guarantees on the physical clock synchronisation of the network, neither of which are required by \acl{TT}.

\section{Conclusions}
A deterministic programming model for distributed systems remains elusive, except for the recent federated flavour of Lingua Franca~\cite{lohstroh2024deterministic}. However, even this model relies on physical clock synchronisation, which suffers from scalability issues.
To address this gap, we introduce the Timetide language for distributed systems, which treats communication delay as a first-class citizen based on the logical synchrony approach. This allows the programmer to reason about the system as if it were centralised and synchronous, where the transmission delays of a system are expressed in a fixed number of logical ticks, rather than physical time.

We introduce and formalise the semantics of Timetide and show how it can be used to model distributed systems, also demonstrating a structural translation to the synchronous language Esterel. In doing so, we show that the distributed Timetide model can be verified using conventional approaches. Furthermore, we describe a class of \textit{Logical Synchrony Network Compatible Architectures} which can implement Timetide programs in a distributed setting.

While this work provides the first logically synchronous programming model, our work has some limitations: \change{There is currently no process to automatically map \ac{TT} programs to distributed \acp{LSN}}, and a concrete way to relate the logical \change{world} of the Timetide model to physical time \change{and} environmental inputs is not provided. \change{Additionally, there can be a need for tasks to have variable execution rates or even be enabled/disabled entirely, which \ac{TT} currently does not support.} We will dwell on these limitations in the near future.

\begin{acks}
The authors from the University of Auckland acknowledge the support of the Google grants \textit{Designing Scalable Synchronous Applications over Google bittide} and \textit{Towards a formally verified autonomous vehicle by leveraging the bittide protocol}. They also acknowledge the biweekly meetings with the Google bittide team, namely Tammo Spalink, Sanjay Lall, and Martin Izzard.
\end{acks}

\bibliographystyle{ACM-Reference-Format}
\bibliography{refs}


\begin{thebibliography}{37}


\ifx \showCODEN    \undefined \def \showCODEN     #1{\unskip}     \fi
\ifx \showDOI      \undefined \def \showDOI       #1{#1}\fi
\ifx \showISBNx    \undefined \def \showISBNx     #1{\unskip}     \fi
\ifx \showISBNxiii \undefined \def \showISBNxiii  #1{\unskip}     \fi
\ifx \showISSN     \undefined \def \showISSN      #1{\unskip}     \fi
\ifx \showLCCN     \undefined \def \showLCCN      #1{\unskip}     \fi
\ifx \shownote     \undefined \def \shownote      #1{#1}          \fi
\ifx \showarticletitle \undefined \def \showarticletitle #1{#1}   \fi
\ifx \showURL      \undefined \def \showURL       {\relax}        \fi
\providecommand\bibfield[2]{#2}
\providecommand\bibinfo[2]{#2}
\providecommand\natexlab[1]{#1}
\providecommand\showeprint[2][]{arXiv:#2}

\bibitem[ptp({[n.\,d.]})]%
        {ptp}
 \bibinfo{year}{[n.\,d.]}\natexlab{}.
\newblock \bibinfo{title}{Precision Clock Synchronization Protocol for Networked Measurement and Control Systems}.
\newblock
\newblock
\newblock
\shownote{IEEE standard 2021.9456762}.


\bibitem[Andalam et~al\mbox{.}(2009)]%
        {andalam2009pret}
\bibfield{author}{\bibinfo{person}{Sidharta Andalam}, \bibinfo{person}{Partha Roop}, \bibinfo{person}{Alain Girault}, {and} \bibinfo{person}{Claus Traulsen}.} \bibinfo{year}{2009}\natexlab{}.
\newblock \emph{\bibinfo{title}{PRET-C: A new language for programming precision timed architectures}}.
\newblock \bibinfo{thesistype}{Ph.\,D. Dissertation}. \bibinfo{school}{INRIA}.
\newblock


\bibitem[Attiya and Hamam(2006)]%
        {attiya2006task}
\bibfield{author}{\bibinfo{person}{Gamal Attiya} {and} \bibinfo{person}{Yskandar Hamam}.} \bibinfo{year}{2006}\natexlab{}.
\newblock \showarticletitle{Task allocation for maximizing reliability of distributed systems: A simulated annealing approach}.
\newblock \bibinfo{journal}{\emph{Journal of parallel and Distributed Computing}} \bibinfo{volume}{66}, \bibinfo{number}{10} (\bibinfo{year}{2006}), \bibinfo{pages}{1259--1266}.
\newblock


\bibitem[Baudisch et~al\mbox{.}(2010)]%
        {baudisch2010dependency}
\bibfield{author}{\bibinfo{person}{Daniel Baudisch}, \bibinfo{person}{Jens Brandt}, {and} \bibinfo{person}{Klaus Schneider}.} \bibinfo{year}{2010}\natexlab{}.
\newblock \showarticletitle{Dependency-driven distribution of synchronous programs}. In \bibinfo{booktitle}{\emph{IFIP Working Conference on Distributed and Parallel Embedded Systems}}. Springer, \bibinfo{pages}{169--180}.
\newblock


\bibitem[Benveniste et~al\mbox{.}(2003)]%
        {benveniste2003synchronous}
\bibfield{author}{\bibinfo{person}{Albert Benveniste}, \bibinfo{person}{Paul Caspi}, \bibinfo{person}{Stephen~A Edwards}, \bibinfo{person}{Nicolas Halbwachs}, \bibinfo{person}{Paul Le~Guernic}, {and} \bibinfo{person}{Robert De~Simone}.} \bibinfo{year}{2003}\natexlab{}.
\newblock \showarticletitle{The synchronous languages 12 years later}.
\newblock \bibinfo{journal}{\emph{Proc. IEEE}} \bibinfo{volume}{91}, \bibinfo{number}{1} (\bibinfo{year}{2003}), \bibinfo{pages}{64--83}.
\newblock


\bibitem[Benveniste et~al\mbox{.}(1991)]%
        {benveniste1991synchronous}
\bibfield{author}{\bibinfo{person}{Albert Benveniste}, \bibinfo{person}{Paul Le~Guernic}, {and} \bibinfo{person}{Christian Jacquemot}.} \bibinfo{year}{1991}\natexlab{}.
\newblock \showarticletitle{Synchronous programming with events and relations: the SIGNAL language and its semantics}.
\newblock \bibinfo{journal}{\emph{Science of computer programming}} \bibinfo{volume}{16}, \bibinfo{number}{2} (\bibinfo{year}{1991}), \bibinfo{pages}{103--149}.
\newblock


\bibitem[Berry and Gonthier(1992)]%
        {berry1992esterel}
\bibfield{author}{\bibinfo{person}{G{\'e}rard Berry} {and} \bibinfo{person}{Georges Gonthier}.} \bibinfo{year}{1992}\natexlab{}.
\newblock \showarticletitle{The Esterel synchronous programming language: Design, semantics, implementation}.
\newblock \bibinfo{journal}{\emph{Science of computer programming}} \bibinfo{volume}{19}, \bibinfo{number}{2} (\bibinfo{year}{1992}), \bibinfo{pages}{87--152}.
\newblock


\bibitem[Berry and Sentovich(2001)]%
        {berry2001multiclock}
\bibfield{author}{\bibinfo{person}{G{\'e}rard Berry} {and} \bibinfo{person}{Ellen Sentovich}.} \bibinfo{year}{2001}\natexlab{}.
\newblock \showarticletitle{Multiclock esterel}. In \bibinfo{booktitle}{\emph{Advanced Research Working Conference on Correct Hardware Design and Verification Methods}}. Springer, \bibinfo{pages}{110--125}.
\newblock


\bibitem[Carloni et~al\mbox{.}(2001)]%
        {carloni2001theory}
\bibfield{author}{\bibinfo{person}{Luca~P Carloni}, \bibinfo{person}{Kenneth~L McMillan}, {and} \bibinfo{person}{Alberto~L Sangiovanni-Vincentelli}.} \bibinfo{year}{2001}\natexlab{}.
\newblock \showarticletitle{Theory of latency-insensitive design}.
\newblock \bibinfo{journal}{\emph{IEEE Transactions on computer-aided design of integrated circuits and systems}} \bibinfo{volume}{20}, \bibinfo{number}{9} (\bibinfo{year}{2001}), \bibinfo{pages}{1059--1076}.
\newblock


\bibitem[Cohen et~al\mbox{.}(2006)]%
        {cohen2006n}
\bibfield{author}{\bibinfo{person}{Albert Cohen}, \bibinfo{person}{Marc Duranton}, \bibinfo{person}{Christine Eisenbeis}, \bibinfo{person}{Claire Pagetti}, \bibinfo{person}{Florence Plateau}, {and} \bibinfo{person}{Marc Pouzet}.} \bibinfo{year}{2006}\natexlab{}.
\newblock \showarticletitle{N-synchronous Kahn networks: a relaxed model of synchrony for real-time systems}.
\newblock \bibinfo{journal}{\emph{ACM SIGPLAN Notices}} \bibinfo{volume}{41}, \bibinfo{number}{1} (\bibinfo{year}{2006}), \bibinfo{pages}{180--193}.
\newblock


\bibitem[Corbett et~al\mbox{.}(2013)]%
        {corbett_spanner_2013}
\bibfield{author}{\bibinfo{person}{James~C. Corbett}, \bibinfo{person}{Jeffrey Dean}, \bibinfo{person}{Michael Epstein}, \bibinfo{person}{Andrew Fikes}, \bibinfo{person}{Christopher Frost}, \bibinfo{person}{J.~J. Furman}, \bibinfo{person}{Sanjay Ghemawat}, \bibinfo{person}{Andrey Gubarev}, \bibinfo{person}{Christopher Heiser}, \bibinfo{person}{Peter Hochschild}, \bibinfo{person}{Wilson Hsieh}, \bibinfo{person}{Sebastian Kanthak}, \bibinfo{person}{Eugene Kogan}, \bibinfo{person}{Hongyi Li}, \bibinfo{person}{Alexander Lloyd}, \bibinfo{person}{Sergey Melnik}, \bibinfo{person}{David Mwaura}, \bibinfo{person}{David Nagle}, \bibinfo{person}{Sean Quinlan}, \bibinfo{person}{Rajesh Rao}, \bibinfo{person}{Lindsay Rolig}, \bibinfo{person}{Yasushi Saito}, \bibinfo{person}{Michal Szymaniak}, \bibinfo{person}{Christopher Taylor}, \bibinfo{person}{Ruth Wang}, {and} \bibinfo{person}{Dale Woodford}.} \bibinfo{year}{2013}\natexlab{}.
\newblock \showarticletitle{Spanner: {Google}’s globally distributed database}.
\newblock \bibinfo{journal}{\emph{ACM Transactions on Computer Systems}} \bibinfo{volume}{31}, \bibinfo{number}{3} (\bibinfo{date}{Aug.} \bibinfo{year}{2013}), \bibinfo{pages}{1--22}.
\newblock
\showISSN{0734-2071, 1557-7333}
\urldef\tempurl%
\url{https://doi.org/10.1145/2491245}
\showDOI{\tempurl}


\bibitem[Edwards(2018)]%
        {edwards2018determinism}
\bibfield{author}{\bibinfo{person}{Stephen~A Edwards}.} \bibinfo{year}{2018}\natexlab{}.
\newblock \showarticletitle{On determinism}.
\newblock \bibinfo{journal}{\emph{Principles of Modeling: Essays Dedicated to Edward A. Lee on the Occasion of His 60th Birthday}} (\bibinfo{year}{2018}), \bibinfo{pages}{240--253}.
\newblock


\bibitem[Ernst et~al\mbox{.}(2018)]%
        {ernst2018system}
\bibfield{author}{\bibinfo{person}{Rolf Ernst}, \bibinfo{person}{Leonie Ahrendts}, {and} \bibinfo{person}{Kai-Bj{\"o}rn Gemlau}.} \bibinfo{year}{2018}\natexlab{}.
\newblock \showarticletitle{System level LET: Mastering cause-effect chains in distributed systems}. In \bibinfo{booktitle}{\emph{IECON 2018-44th Annual Conference of the IEEE Industrial Electronics Society}}. IEEE, \bibinfo{pages}{4084--4089}.
\newblock


\bibitem[Floudas and Lin(2005)]%
        {floudas2005mixed}
\bibfield{author}{\bibinfo{person}{Christodoulos~A Floudas} {and} \bibinfo{person}{Xiaoxia Lin}.} \bibinfo{year}{2005}\natexlab{}.
\newblock \showarticletitle{Mixed integer linear programming in process scheduling: Modeling, algorithms, and applications}.
\newblock \bibinfo{journal}{\emph{Annals of Operations Research}}  \bibinfo{volume}{139} (\bibinfo{year}{2005}), \bibinfo{pages}{131--162}.
\newblock


\bibitem[Gilles(1974)]%
        {gilles1974semantics}
\bibfield{author}{\bibinfo{person}{KAHN Gilles}.} \bibinfo{year}{1974}\natexlab{}.
\newblock \showarticletitle{The semantics of a simple language for parallel programming}.
\newblock \bibinfo{journal}{\emph{Information processing}} \bibinfo{volume}{74}, \bibinfo{number}{471-475} (\bibinfo{year}{1974}), \bibinfo{pages}{15--28}.
\newblock


\bibitem[Girault({[n.\,d.]})]%
        {girault2005survey}
\bibfield{author}{\bibinfo{person}{Alain Girault}.} \bibinfo{year}{[n.\,d.]}\natexlab{}.
\newblock \showarticletitle{A survey of automatic distribution method for synchronous programs}.
\newblock


\bibitem[G{\"u}nzel et~al\mbox{.}(2021)]%
        {gunzel2021timing}
\bibfield{author}{\bibinfo{person}{Mario G{\"u}nzel}, \bibinfo{person}{Kuan-Hsun Chen}, \bibinfo{person}{Niklas Ueter}, \bibinfo{person}{Georg von~der Br{\"u}ggen}, \bibinfo{person}{Marco D{\"u}rr}, {and} \bibinfo{person}{Jian-Jia Chen}.} \bibinfo{year}{2021}\natexlab{}.
\newblock \showarticletitle{Timing analysis of asynchronized distributed cause-effect chains}. In \bibinfo{booktitle}{\emph{2021 IEEE 27th Real-Time and Embedded Technology and Applications Symposium (RTAS)}}. IEEE, \bibinfo{pages}{40--52}.
\newblock


\bibitem[Halbwachs et~al\mbox{.}(1991)]%
        {halbwachs1991synchronous}
\bibfield{author}{\bibinfo{person}{Nicolas Halbwachs}, \bibinfo{person}{Paul Caspi}, \bibinfo{person}{Pascal Raymond}, {and} \bibinfo{person}{Daniel Pilaud}.} \bibinfo{year}{1991}\natexlab{}.
\newblock \showarticletitle{The synchronous data flow programming language LUSTRE}.
\newblock \bibinfo{journal}{\emph{Proc. IEEE}} \bibinfo{volume}{79}, \bibinfo{number}{9} (\bibinfo{year}{1991}), \bibinfo{pages}{1305--1320}.
\newblock


\bibitem[Henzinger et~al\mbox{.}(2001)]%
        {henzinger2001giotto}
\bibfield{author}{\bibinfo{person}{Thomas~A Henzinger}, \bibinfo{person}{Benjamin Horowitz}, {and} \bibinfo{person}{Christoph~Meyer Kirsch}.} \bibinfo{year}{2001}\natexlab{}.
\newblock \showarticletitle{Giotto: A time-triggered language for embedded programming}. In \bibinfo{booktitle}{\emph{Embedded Software: First International Workshop, EMSOFT 2001 Tahoe City, CA, USA, October 8--10, 2001 Proceedings 1}}. Springer, \bibinfo{pages}{166--184}.
\newblock


\bibitem[Hoare et~al\mbox{.}(1985)]%
        {hoare1985communicating}
\bibfield{author}{\bibinfo{person}{Charles Antony~Richard Hoare} {et~al\mbox{.}}} \bibinfo{year}{1985}\natexlab{}.
\newblock \bibinfo{booktitle}{\emph{Communicating sequential processes}}. Vol.~\bibinfo{volume}{178}.
\newblock \bibinfo{publisher}{Prentice-hall Englewood Cliffs}.
\newblock


\bibitem[Kenwright et~al\mbox{.}(2024)]%
        {kenwright2024logical}
\bibfield{author}{\bibinfo{person}{Logan Kenwright}, \bibinfo{person}{Partha Roop}, \bibinfo{person}{Nathan Allen}, \bibinfo{person}{Sanjay Lall}, \bibinfo{person}{C{\u{a}}lin Ca{\c{s}}caval}, \bibinfo{person}{Tammo Spalink}, {and} \bibinfo{person}{Martin Izzard}.} \bibinfo{year}{2024}\natexlab{}.
\newblock \showarticletitle{Logical Synchrony Networks: A formal model for deterministic distribution}.
\newblock \bibinfo{journal}{\emph{IEEE Access}} (\bibinfo{year}{2024}).
\newblock


\bibitem[Kirsch and Sokolova(2012)]%
        {kirsch2012logical}
\bibfield{author}{\bibinfo{person}{Christoph~M Kirsch} {and} \bibinfo{person}{Ana Sokolova}.} \bibinfo{year}{2012}\natexlab{}.
\newblock \showarticletitle{The logical execution time paradigm}.
\newblock \bibinfo{journal}{\emph{Advances in Real-Time Systems}} (\bibinfo{year}{2012}), \bibinfo{pages}{103--120}.
\newblock


\bibitem[Kroening and Tautschnig(2014)]%
        {kroening2014cbmc}
\bibfield{author}{\bibinfo{person}{Daniel Kroening} {and} \bibinfo{person}{Michael Tautschnig}.} \bibinfo{year}{2014}\natexlab{}.
\newblock \showarticletitle{CBMC--C Bounded Model Checker: (Competition Contribution)}. In \bibinfo{booktitle}{\emph{Tools and Algorithms for the Construction and Analysis of Systems: 20th International Conference, TACAS 2014, Held as Part of the European Joint Conferences on Theory and Practice of Software, ETAPS 2014, Grenoble, France, April 5-13, 2014. Proceedings 20}}. Springer, \bibinfo{pages}{389--391}.
\newblock


\bibitem[Lall et~al\mbox{.}(2024)]%
        {lall_logical_synchrony_2024}
\bibfield{author}{\bibinfo{person}{Sanjay Lall}, \bibinfo{person}{C\u{a}lin Ca\c{s}caval}, \bibinfo{person}{Martin Izzard}, {and} \bibinfo{person}{Tammo Spalink}.} \bibinfo{year}{2024}\natexlab{}.
\newblock \showarticletitle{Logical {Synchrony} and the bittide {Mechanism}}.
\newblock \bibinfo{journal}{\emph{IEEE Transactions on Parallel and Distributed Systems}} (\bibinfo{year}{2024}), \bibinfo{pages}{1--14}.
\newblock
\showISSN{1045-9219, 1558-2183, 2161-9883}
\urldef\tempurl%
\url{https://doi.org/10.1109/TPDS.2024.3444739}
\showDOI{\tempurl}


\bibitem[Lall et~al\mbox{.}(2022)]%
        {lall2022modeling}
\bibfield{author}{\bibinfo{person}{Sanjay Lall}, \bibinfo{person}{C{\u{a}}lin Ca{\c{s}}caval}, \bibinfo{person}{Martin Izzard}, {and} \bibinfo{person}{Tammo Spalink}.} \bibinfo{year}{2022}\natexlab{}.
\newblock \showarticletitle{Modeling and Control of bittide Synchronization}. In \bibinfo{booktitle}{\emph{2022 American Control Conference (ACC)}}. IEEE, \bibinfo{pages}{5185--5192}.
\newblock


\bibitem[Lee(2021)]%
        {lee2021determinism}
\bibfield{author}{\bibinfo{person}{Edward~A Lee}.} \bibinfo{year}{2021}\natexlab{}.
\newblock \showarticletitle{Determinism}.
\newblock \bibinfo{journal}{\emph{ACM Transactions on Embedded Computing Systems (TECS)}} \bibinfo{volume}{20}, \bibinfo{number}{5} (\bibinfo{year}{2021}), \bibinfo{pages}{1--34}.
\newblock


\bibitem[Lee et~al\mbox{.}(2023)]%
        {lee2023consistency}
\bibfield{author}{\bibinfo{person}{Edward~A Lee}, \bibinfo{person}{Ravi Akella}, \bibinfo{person}{Soroush Bateni}, \bibinfo{person}{Shaokai Lin}, \bibinfo{person}{Marten Lohstroh}, {and} \bibinfo{person}{Christian Menard}.} \bibinfo{year}{2023}\natexlab{}.
\newblock \showarticletitle{Consistency vs. availability in distributed cyber-physical systems}.
\newblock \bibinfo{journal}{\emph{ACM Transactions on Embedded Computing Systems}} \bibinfo{volume}{22}, \bibinfo{number}{5s} (\bibinfo{year}{2023}), \bibinfo{pages}{1--24}.
\newblock


\bibitem[Li et~al\mbox{.}(2020)]%
        {li_sundial_2020}
\bibfield{author}{\bibinfo{person}{Yuliang Li} {et~al\mbox{.}}} \bibinfo{year}{2020}\natexlab{}.
\newblock \showarticletitle{Sundial: fault-tolerant clock synchronization for datacenters}. In \bibinfo{booktitle}{\emph{14th {USENIX} Symposium on Operating Systems Design and Implementation ({OSDI} 20)}}. \bibinfo{pages}{1171--1186}.
\newblock


\bibitem[Lohstroh et~al\mbox{.}(2024)]%
        {lohstroh2024deterministic}
\bibfield{author}{\bibinfo{person}{Marten Lohstroh}, \bibinfo{person}{Soroush Bateni}, \bibinfo{person}{Christian Menard}, \bibinfo{person}{Alexander Schulz-Rosengarten}, \bibinfo{person}{Jeronimo Castrillon}, {and} \bibinfo{person}{Edward~A Lee}.} \bibinfo{year}{2024}\natexlab{}.
\newblock \showarticletitle{Deterministic coordination across multiple timelines}.
\newblock \bibinfo{journal}{\emph{ACM Transactions on Embedded Computing Systems}} \bibinfo{volume}{23}, \bibinfo{number}{5} (\bibinfo{year}{2024}), \bibinfo{pages}{1--29}.
\newblock


\bibitem[Lohstroh et~al\mbox{.}(2021)]%
        {lohstroh2021toward}
\bibfield{author}{\bibinfo{person}{Marten Lohstroh}, \bibinfo{person}{Christian Menard}, \bibinfo{person}{Soroush Bateni}, {and} \bibinfo{person}{Edward~A Lee}.} \bibinfo{year}{2021}\natexlab{}.
\newblock \showarticletitle{Toward a lingua franca for deterministic concurrent systems}.
\newblock \bibinfo{journal}{\emph{ACM Transactions on Embedded Computing Systems (TECS)}} \bibinfo{volume}{20}, \bibinfo{number}{4} (\bibinfo{year}{2021}), \bibinfo{pages}{1--27}.
\newblock


\bibitem[Mandel et~al\mbox{.}(2010)]%
        {mandel2010lucy}
\bibfield{author}{\bibinfo{person}{Louis Mandel}, \bibinfo{person}{Florence Plateau}, {and} \bibinfo{person}{Marc Pouzet}.} \bibinfo{year}{2010}\natexlab{}.
\newblock \showarticletitle{Lucy-n: a n-synchronous extension of Lustre}. In \bibinfo{booktitle}{\emph{Mathematics of Program Construction: 10th International Conference, MPC 2010, Qu{\'e}bec City, Canada, June 21-23, 2010. Proceedings 10}}. Springer, \bibinfo{pages}{288--309}.
\newblock


\bibitem[Milner(1984)]%
        {milner1984lectures}
\bibfield{author}{\bibinfo{person}{Robin Milner}.} \bibinfo{year}{1984}\natexlab{}.
\newblock \showarticletitle{Lectures on a calculus for communicating systems}. In \bibinfo{booktitle}{\emph{International Conference on Concurrency}}. Springer, \bibinfo{pages}{197--220}.
\newblock


\bibitem[Plotkin(2004)]%
        {plotkin2004origins}
\bibfield{author}{\bibinfo{person}{Gordon~D Plotkin}.} \bibinfo{year}{2004}\natexlab{}.
\newblock \showarticletitle{The origins of structural operational semantics}.
\newblock \bibinfo{journal}{\emph{The Journal of Logic and Algebraic Programming}}  \bibinfo{volume}{60} (\bibinfo{year}{2004}), \bibinfo{pages}{3--15}.
\newblock


\bibitem[Siron et~al\mbox{.}(2022)]%
        {siron2022synchronous}
\bibfield{author}{\bibinfo{person}{Fabien Siron}, \bibinfo{person}{Dumitru Potop-Butucaru}, \bibinfo{person}{Robert De~Simone}, \bibinfo{person}{Damien Chabrol}, {and} \bibinfo{person}{Amira Methni}.} \bibinfo{year}{2022}\natexlab{}.
\newblock \showarticletitle{The synchronous Logical Execution Time paradigm}. In \bibinfo{booktitle}{\emph{ERTS 2022-Embedded real time systems}}.
\newblock


\bibitem[Siron et~al\mbox{.}(2023)]%
        {siron2023formal}
\bibfield{author}{\bibinfo{person}{Fabien Siron}, \bibinfo{person}{Dumitru Potop-Butucaru}, \bibinfo{person}{Robert De~Simone}, \bibinfo{person}{Damien Chabrol}, {and} \bibinfo{person}{Amira Methni}.} \bibinfo{year}{2023}\natexlab{}.
\newblock \emph{\bibinfo{title}{Formal Semantics of the PsyC language}}.
\newblock \bibinfo{thesistype}{Ph.\,D. Dissertation}. \bibinfo{school}{Inria-Sophia Antipolis}.
\newblock


\bibitem[Treat(2025)]%
        {comcast_tool}
\bibfield{author}{\bibinfo{person}{Tyler Treat}.} \bibinfo{year}{2025}\natexlab{}.
\newblock \bibinfo{title}{comcast}.
\newblock
\newblock
\urldef\tempurl%
\url{https://github.com/tylertreat/comcast}
\showURL{%
\tempurl}
\newblock
\shownote{original-date: 2014-11-12}.


\bibitem[Tripakis et~al\mbox{.}(2008)]%
        {tripakis2008implementing}
\bibfield{author}{\bibinfo{person}{Stavros Tripakis}, \bibinfo{person}{Claudio Pinello}, \bibinfo{person}{Albert Benveniste}, \bibinfo{person}{Alberto Sangiovanni-Vincent}, \bibinfo{person}{Paul Caspi}, {and} \bibinfo{person}{Marco Di~Natale}.} \bibinfo{year}{2008}\natexlab{}.
\newblock \showarticletitle{Implementing synchronous models on loosely time triggered architectures}.
\newblock \bibinfo{journal}{\emph{IEEE Trans. Comput.}} \bibinfo{volume}{57}, \bibinfo{number}{10} (\bibinfo{year}{2008}), \bibinfo{pages}{1300--1314}.
\newblock


\end{thebibliography}


\end{document}